\documentclass[%
 aip,
 jmp,%
 amsmath,amssymb,
 reprint,%
 nofootinbib,
]{revtex4-2}

\usepackage{algorithm,algpseudocode}

\usepackage{xcolor}
\usepackage{graphicx}% Include figure files
\usepackage{dcolumn}% Align table columns on decimal point
\usepackage{bbm}

\newcommand{\nobtained}{$\mathsf{Number}$ $\mathsf{Obtained}$}
\newcommand{\Jposterior}{$\mathsf{Posterior}$ $\mathsf{Jaccard}$ $\mathsf{Index}$}
\newcommand{\GP}{$\mathcal{GP}$}
\newcommand{\posteriormean}{$\bar{f}$}
\newcommand{\posteriorsamples}{$\{f_i \}_{i = 1}^n$}
\newcommand{\alg}{$\mathcal{A}$}

\newcommand{\SIysampling}{S1}
\newcommand{\SIxsampling}{S2}
\newcommand{\SInoise}{S3}
\newcommand{\SIconditional}{S4}
\newcommand{\SIpercentile}{S5}
\newcommand{\SIfixedhypers}{S6}
\newcommand{\SIpareto}{S7}

\let\svthefootnote\thefootnote
\newcommand\freefootnote[1]{%
  \let\thefootnote\relax%
  \footnotetext{#1}%
  \let\thefootnote\svthefootnote%
}

\usepackage{natbib}
\setcitestyle{numbers}
\setcitestyle{square}

\begin{document}

% \setcitestyle{square,numbers,sort&compress}

\preprint{AIP/123-QED}

\title{\LARGE \centering Targeted materials discovery using \\Bayesian algorithm execution 
\vspace{5mm}}% Force line breaks with \\

\author{Sathya R. Chitturi}
\thanks{Email: chitturi@stanford.edu}
\affiliation{%
  \small
  SLAC National Accelerator Laboratory, Menlo Park, CA 94025
}
\affiliation{%
  \small
  \mbox{Department of Materials Science and Engineering, Stanford University, Stanford,}\\\hspace{2mm}CA 94305
}

\author{Akash Ramdas}
\affiliation{%
  \small
  SLAC National Accelerator Laboratory, Menlo Park, CA 94025
}
\affiliation{%
  \small
  \mbox{Department of Materials Science and Engineering, Stanford University, Stanford,}\\\hspace{2mm}CA 94305
}

\author{Yue Wu}
\affiliation{%
  \small
  SLAC National Accelerator Laboratory, Menlo Park, CA 94025
}

\author{Brian Rohr}
\affiliation{%
  \small
  SLAC National Accelerator Laboratory, Menlo Park, CA 94025
}

\author{Stefano Ermon}
\affiliation{%
  \small
  \mbox{Computer Science Department, Stanford University, Stanford, CA 94305}
}

\author{Jennifer Dionne}
\affiliation{%
  \small
  \mbox{Department of Materials Science and Engineering, Stanford University, Stanford,}\\\hspace{2mm}CA 94305
}

\author{Felipe H. da Jornada}
\affiliation{%
  \small
  SLAC National Accelerator Laboratory, Menlo Park, CA 94025
}
\affiliation{%
  \small
  \mbox{Department of Materials Science and Engineering, Stanford University, Stanford,}\\\hspace{2mm}CA 94305
}

\author{Mike Dunne}
\affiliation{%
  \small
  SLAC National Accelerator Laboratory, Menlo Park, CA 94025
}

\author{Christopher Tassone}
\thanks{These authors co-supervised the work}
\affiliation{%
  \small
  SLAC National Accelerator Laboratory, Menlo Park, CA 94025
}

\author{Willie~Neiswanger}
\thanks{These authors co-supervised the work}
\affiliation{
  \small
  \mbox{Department of Computer Science, University of Southern California, Los Angeles,}\\\hspace{2mm}CA 90089
}
\affiliation{%
  \small
  \mbox{Computer Science Department, Stanford University, Stanford, CA 94305}
}

\author{Daniel Ratner\vspace{5mm}}
\thanks{These authors co-supervised the work}
\affiliation{%
  \small
  SLAC National Accelerator Laboratory, Menlo Park, CA 94025
}

% \freefootnote{* Email: chitturi@stanford.edu}
% \freefootnote{C.J.T, W.N, and D.R co-supervised this work.}

% \date{\today}% It is always \today, today,
             %  but any date may be explicitly specified

\begin{abstract}
\vspace{2mm}
\textbf{\textsf{Abstract}}.
Rapid discovery and synthesis of new materials requires intelligent data acquisition strategies to navigate large design spaces. A popular strategy is Bayesian optimization, which aims to find candidates that maximize material properties; however, materials design often requires finding specific subsets of the design space which meet more complex or specialized goals. We present a framework that captures experimental goals through straightforward user-defined filtering algorithms. These algorithms are automatically translated into one of three intelligent, parameter-free, sequential data acquisition strategies (SwitchBAX, InfoBAX, and MeanBAX). Our framework is tailored for typical discrete search spaces involving multiple measured physical properties and short time-horizon decision making. We evaluate this approach on datasets for TiO$_\text{2}$ nanoparticle synthesis and magnetic materials characterization, and show that our methods are significantly more efficient than state-of-the-art approaches.
\end{abstract}

\keywords{Autonomous Experimentation, Design of Experiments, Machine Learning, Materials Science, Materials Discovery, Bayesian Optimization \vspace{-3mm}}%Use showkeys class option if keyword
                              %display desired

\maketitle

\section*{\label{sec:level1}Introduction}
Modern materials discovery involves searching large regions of multi-dimensional processing or synthesis conditions to find candidate materials that achieve specific desired properties. For example, the lithium-ion batteries that have enabled both the personal electronics and clean mobility revolutions started out using simple LiCoO\(_2\) as the cathode active material, but this has given way to numerous formulations of the form Li(Ni$_\text{1/3+x}$Co$_{\text{1/3-2x}}$Mn$_\text{{1/3+x}}$)O\(_2\) where each metal contributes to various aspects of stability and electrochemistry \cite{goodenough2013li}. Another example is in the development of high temperature superconducting materials where the trade-off between different quantum phenomena (e.g., charge density waves and the superconducting state) needs to be addressed via iterative synthesis and characterization of tailored materials \cite{lee2006doping}. Often, the rate of discovery is naturally limited by the speed at which experiments can be performed; this is particularly true for materials applications involving low levels of automation, complex multi-step synthesis protocols, and slow/expensive characterization modalities. For these important situations, developing algorithms which can quickly identify desirable conditions under limited experimental budgets is critical to furthering materials discovery \cite{suh2020evolving,montoya2022toward}. 

Intelligent sequential experimental design has emerged as a promising approach to rapidly search large design spaces. Compared to classical techniques such as factorial design of experiments, sequential methods use new data collected at each step to reduce the total number of experiments needed to find optimal designs \cite{shahriari2015taking,kusne2020fly}. Current methods typically involve two components: 1) a probabilistic statistical model trained to predict both the value and the uncertainty of a measurable property at any point in the design space (here, defined as a discrete set of all possible measurement or synthesis conditions) and 2) an `acquisition function' which assigns a relative numerical score to each point in the design space. Under this paradigm, measurements are made at the design point which has the highest acquisition value. 

\begin{figure}
\centering
\includegraphics[width=0.9\linewidth]{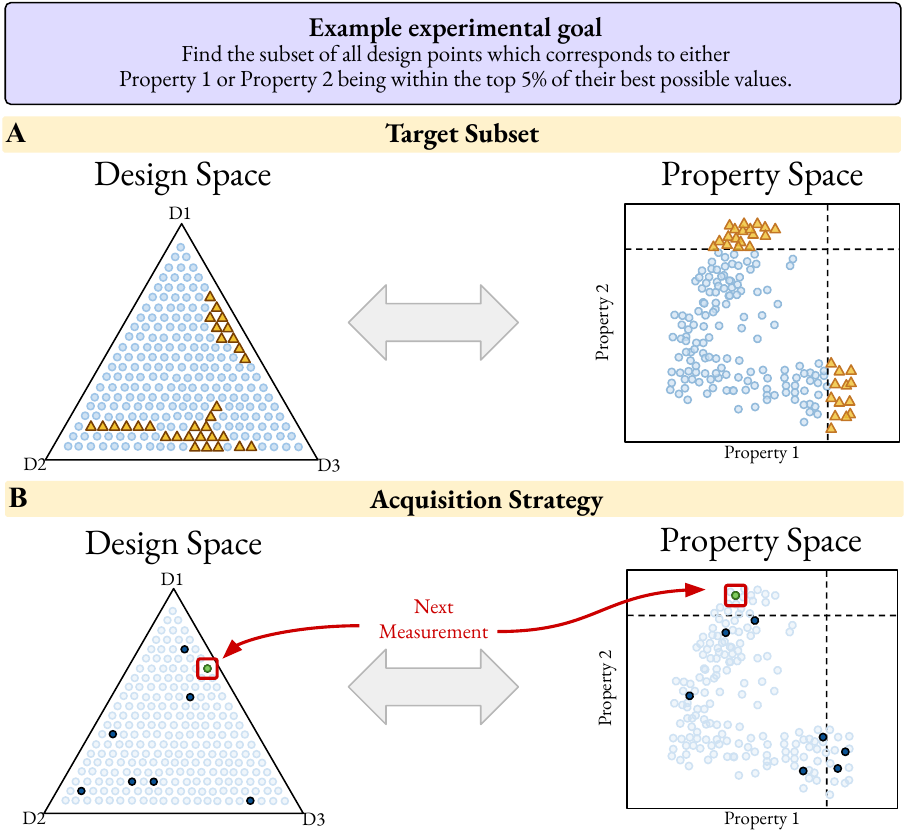}
\caption{Specification of an example experimental goal and translation into an automated data-acquisition strategy. (\textbf{A}) Visualization of the design space and corresponding measurement space for an example materials system. Samples from the design space (a discrete set of design points) map directly to a set of measured properties (measured property space). The set of all possible design points and measurable properties are shown in blue. The ground-truth target subset of the design space corresponding to the user-goal is shown in orange. Importantly, the ground-truth subset which achieves the experimental goal is unknown prior to experimentation. (\textbf{B}) New data is acquired intelligently based on both previously collected measurements \textbf{and} the specific experimental goal. The method for achieving this recommendation strategy is the focus of the manuscript.}
\label{fig:overall}
\end{figure}

No matter the accuracy of the model, intelligent data acquisition strategies will be limited by the \textit{relevance} of the acquisition function, i.e. how closely the acquisition function aligns with the user's experimental goal. In this work, we focus on the problem of \textbf{automatically creating custom acquisition functions to target specific experimental goals}. This is an important problem, as materials applications often involve precise requirements that are not well addressed by existing sequential design of experiment techniques. Specifically, we will consider the task of finding the `target subset' of the design space that satisfies user-defined criteria on measured properties. An example of a custom experimental goal, the corresponding target subset of the design space and data acquisition scheme is shown in Figure \ref{fig:overall}.

Most prior work in adaptive decision making has focused on the goal of single objective optimization: finding the design point corresponding to the global optimum for a property of interest \cite{kochenderfer2019algorithms}. An example of this type of goal is the task of developing novel electrolyte formulations with the largest electrochemical windows of stability \cite{dave2020autonomous}. For single objective optimization, the framework of Bayesian optimization (BO) applies, and has a variety of relevant acquisition functions including Upper Confidence Bound (UCB), Probability of Improvement (PI), and Expected Improvement (EI) \cite{greenhill2020bayesian,kochenderfer2019algorithms}. For multi-property optimization, typically there does not exist a single design condition that is optimal with respect to all properties. Instead, the goal is to obtain the set of design points which optimally trade-off between objectives (Pareto optimal designs) \cite{greenhill2020bayesian}. Common multi-objective Bayesian optimization acquisition functions include Expected Hypervolume Improvement (EHVI) \cite{emmerich2011hypervolume,daulton2020differentiable}, Noisy Hypervolume Improvement (NEHVI) \cite{daulton2021parallel}, and ParEGO \cite{knowles2006parego}. Single and multi-objective Bayesian optimization have been applied in a number of materials settings \cite{dave2020autonomous,wang2022benchmarking,hase2018phoenics,rohr2020benchmarking,yamashita2018crystal,hickman2022bayesian,herbol2018efficient,zhang2020bayesian,liang2021benchmarking,shields2021bayesian,hase2021gryffin,hanaoka2021bayesian,karasuyama2020computational,hu2023multi,khatamsaz2023bayesian,xu2023bayesian,hu2023multi,wang2023bayesian}. For further details on materials-focused Bayesian optimization,  see \cite{packwood2017bayesian}.  

Another well-studied experimental goal is mapping (full-function estimation). Instead of finding global optima, the task is instead to learn the relationship between the design space and the property space. Uncertainty Sampling (US) is a typical acquisition function for this purpose. Such strategies have been used to achieve higher image resolutions in shorter collection times and have found application in fields such as X-ray scattering \cite{yager2023autonomous,noack2019kriging,szymanski2023adaptively} and microscopy \cite{kalinin2021automated,ament2021autonomous}. Generally, mapping tasks are useful in helping elucidate insights about the entire system but come with the downside of needing to perform a large number of (potentially slow) experiments across the entire design space. 

The primary subject of this manuscript addresses the larger goal of \textbf{finding specific target regions of the design space which conform to specific conditions on the properties}, which subsumes the aforementioned goals of optimization and full-function estimation, as well as other more complex tasks including level-set estimation \cite{bogunovic2016truncated,ha2021high}. In this more general setting, the goal is to isolate the set of specific design points which achieve precise user-specified property criteria. For subsets that do not involve optimization or mapping, either custom acquisition functions need to be developed or users are forced to use existing acquisition functions which are not necessarily aligned (and thus inefficient) for their specific experimental task. Developing new acquisition functions is possible \cite{terayama2019efficient,dai2020efficient,kusne2020fly}, but this often requires significant time and mathematical insight. These limitations restrict accessibility to the broader materials community and hinders the pace of materials innovation.

Various important scientific problems fall into the category of subset estimation, including: determining synthesis conditions targeting varying ranges of monodisperse colloidal nanoparticle sizes for heterogeneous catalysis \cite{fong2021utilization} or plasmonics \cite{feng2020investigation}, enumerating processing conditions corresponding to wide stability windows \cite{rivnay2016structural}, accurately mapping specific portions of phase boundaries \cite{terayama2019efficient,dai2020efficient,kusne2020fly}, charting transition state pathways between distant structural minima in a potential energy landscape \cite{prentiss2010energy, singh2018computational}, and finding chemically diverse sets of ligands that are strong, non-toxic binders \cite{foloppe2006identification}. The ability to obtain \textit{sets} of design points which meet user-specifications is particularly important from a practical adoption standpoint. Many novel materials do not achieve widespread industrial application due to long-term failure modes. Common failure modes include degradation mechanisms in batteries \cite{palacin2016batteries}, catalysts \cite{scott2018matter}, and solar cells \cite{jorgensen2008stability}, and toxicity of various bio-compatible materials  and medical therapeutics \cite{di2015drug}. Obtaining a large pool of plausible designs can mitigate against the risk of long term failure, improving the odds of discovering transformative materials. It is worth mentioning that these problems involve identifying a larger set of design points than optimization (which is typically only a few design points) and a substantially smaller region than full-function estimation (the entire domain). Note, while multi-objective optimization does aim to locate a set of design points, this procedure only returns a specific set called the Pareto front, corresponding to the optimal trade-off between measured properties.

In this manuscript, we present a framework for building acquisition functions that can precisely target a subset of the design space corresponding to an experimental goal. The user defines their goal via an algorithmic procedure that would return the correct subset of the design space if the underlying mapping were known. This algorithm undergoes an automatic conversion into an acquisition function that can guide future experimentation, bypassing the need to devise complex acquisition functions for specific applications.

Our work presents both methodological development and showcases first application to the domain of materials research. Specifically, we adapt information-based Bayesian Algorithm Execution (InfoBAX) \cite{neiswanger2021bayesian} and Multipoint-BAX \cite{miskovich2022bayesian} to handle materials science scenarios, characterized by discrete design spaces and multi-property measurements. Second, we develop a multi-property generalization of an exploration strategy that uses model posteriors \cite{katsube2020experimental,torres2019low,tian2021determining,dai2020efficient}, which we call MeanBAX. We observe that MeanBAX and InfoBAX exhibit complementary performance in the small-data and medium-data regimes, respectively. For this reason, we additionally design a parameter-free strategy, named SwitchBAX, which is able to dynamically switch between InfoBAX and MeanBAX, that performs well across the full dataset size range. 

For all three approaches, we provide scientists with a simple \href{https://github.com/src47/multibax-sklearn}{open-source interface} to cleanly and simply express complex experimental goals, implement a variety of custom user-defined algorithms tailored to materials estimation problems, and significantly, evaluate the suitability of the BAX framework to guide practical materials experiments. We highlight the applicability of the multi-property BAX strategies by targeting a series of user-defined regions in two datasets from the domains of nanomaterials synthesis and high-throughput magnetic materials characterization. We anticipate that this method will enable the ability to target important non-trivial experimental goals, paving the way for the accelerated design of advanced materials.

\section*{Sequential Experimentation Design Approach}

\subsection*{Expressing an Experimental Goal via Algorithm Execution} 

\begin{figure*}[t!] % this float will be placed at top of p. 2
\centering
\includegraphics[width=\linewidth]{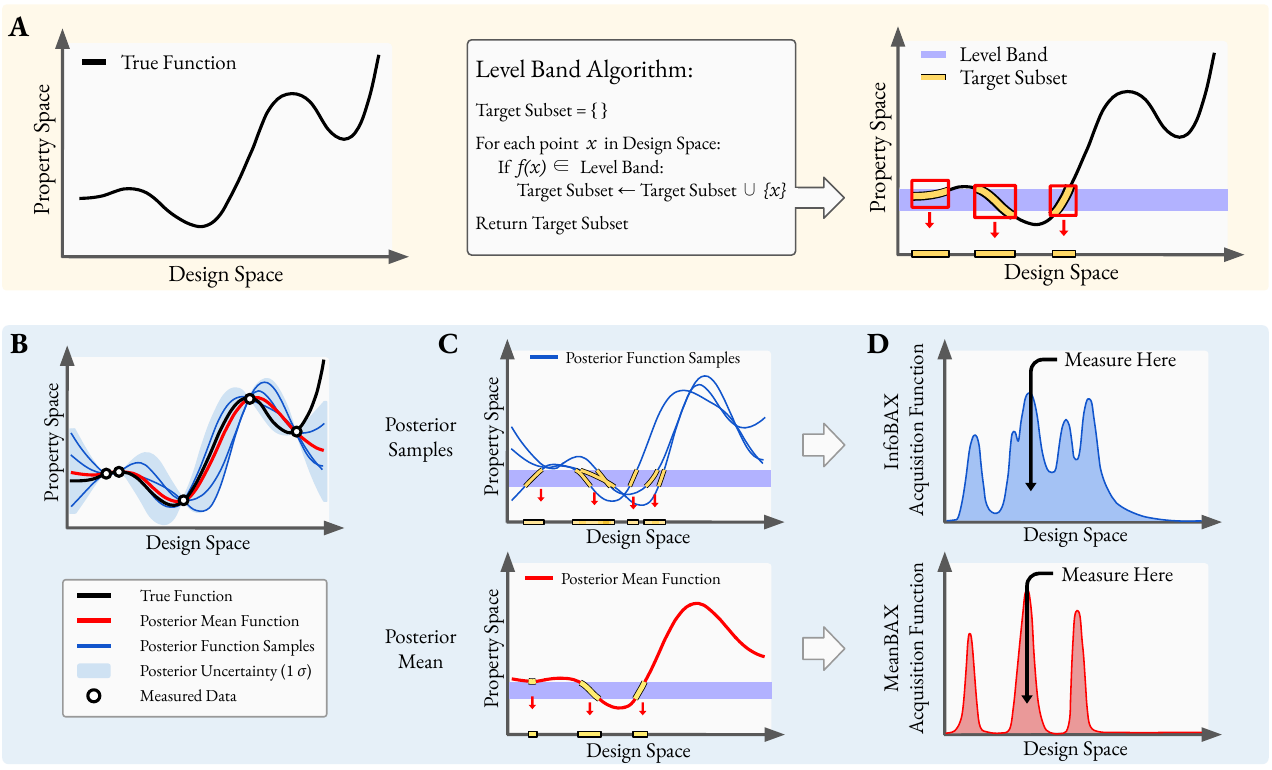}
\caption{(\textbf{A}) Example of a user specified algorithm (Level Band) executed on a true and unknown measured property. Here, the target subset of the design space are the specific set of design points which have measured property values which fall within the specified level band. (\textbf{B}) An illustration of using a Gaussian Process model (\GP{}), a model that predicts a mean value (red line) and an uncertainty (blue band) for every point in the design space, to fit measured data sampled from the true function. (\textbf{C}) Posterior function samples (\posteriorsamples{}) are obtained from this probabilistic model via sampling and represent statistically consistent guesses of the true function based on measured training data. The user algorithm can be executed on either the posterior samples or the posterior mean to build the (\textbf{D}) BAX acquisition functions (InfoBAX and MeanBAX). The next suggested point to measure corresponds to the design point with the highest acquisition value.}
\label{fig:approach_overview}
\end{figure*}

We first consider a design space: a discrete set of $N$ possible synthesis or measurement conditions, each with dimensionality $d$ corresponding to different changeable parameters. Here, $X \in \mathbb{R}^{N \times d}$ is the discrete design space and $x \in \mathbb{R}^{d}$ is a single point in the design space with $d$ features. For each design point, it is possible to perform a costly or time-consuming experiment to obtain a set of $m$ measured properties ($y \in \mathbb{R}^{m}$). The total set of measured properties (measured property space) across the entire design space is denoted $Y \in \mathbb{R}^{N \times m}$. The design space ($X$) and corresponding measurement space ($Y$) are linked through some true noiseless underlying function, $f_*$ which is assumed to be unknown (or black-box) prior to any experimentation (Equation \ref{eqn:mapping}). Real measurements have an additional term, $\epsilon$, corresponding to `measurement noise', which we assume can be modeled by independent and identically distributed normal random variables:
\begin{equation}
    y = f_*(x) + \epsilon \hspace{1mm},\quad \epsilon \sim \mathcal{N}(0, \sigma^2).
    \label{eqn:mapping}
\end{equation}
Within the full design space, there are often specific portions which are particularly desirable to measure. For the purposes of this manuscript, achieving a custom experimental goal is equivalent to finding a specific ground-truth target subset of the design space. We define the ground-truth target subset as $\mathcal{T}_* = \{ \mathcal{T}_*^x, f_*(\mathcal{T}_*^x) \}$. Here, $\mathcal{T}^x_*$ corresponds to the design points which achieve the experimental goal ($\mathcal{T}^x_* \subseteq X$) and $f_*(\mathcal{T}_*^x)$ corresponds to the corresponding underlying property values. Note, in this framework, the experimental goal dictates the underlying target subset. As an example, one specific goal is that of single-property optimization. Here, $\mathcal{T}^x_*$ would refer to the point (or degenerate sets of points) in the design space with the optimal property value; this is the setting of classical Bayesian optimization. However, the subset can also be more complex. In Figure \ref{fig:approach_overview}A, we consider a simple experiment of a single property and single design feature (one-dimensional $Y$ and $X$). In this scenario, the experimental goal is to find the set of points in the design space for which the material property falls within a band between two specified property value thresholds. 

Having defined the ground-truth target subset, we now turn to the concept of defining this subset via an algorithm. First, let us assume that $f_*$ is known throughout the design space. If this were the case, we could \textbf{execute an algorithmic procedure} (\alg{}) to obtain the region, $\mathcal{T}_*$, of interest as
\begin{equation*}
    \mathcal{A}(f_*, X) \rightarrow \mathcal{T}_* .
\label{eqn:algtruefn}
\end{equation*}
Figure \ref{fig:approach_overview}A shows the correspondence between the experimental goal and target subset using the Level Band algorithm. Here the algorithm $\mathcal{A}$ simply scans through every point in the design space and returns the subset for which the property value falls within the level band. 

Of course, the catch is that clearly the underlying mapping, $f_*$, is unknown. However, it turns out that framing an experimental goal as an algorithm that would correctly yield the ground-truth target subset \textit{if the true mapping were known} is a powerful concept. It allows the user to precisely state their desired outcome and in the next section, we will see how to sequentially acquire data to \textit{estimate} the result of the algorithm running on $f_*$.

\subsection*{Obtaining a Target Subset Using BAX}

Bayesian algorithm execution (BAX) is the idea that one may instead execute an algorithm on approximate fitting models (surrogate models) that are designed to mimic the true function but are trained on measurements at only a small number design points. Unlike $f_*$, these models are fast and inexpensive to evaluate. Data is then acquired in a sequential manner to help estimate the true algorithm output (i.e. the result if the algorithm were to be executed on the true, unknown function). We denote $\mathcal{D}_t = \{x_k, y_k\}_{k=1}^t$ as the measured dataset at iteration $t$; $\mathcal{D}^x_t \in \mathbb{R}^{t \times d}$ and $\mathcal{D}^y_t \in \mathbb{R}^{t \times m}$ refer to the $x$ and $y$ components of the dataset. 

The surrogate models are `probabilistic models' which predict both an average response and an uncertainty estimate for every point in the design space. Specifically, in this work, we use a machine learning model known as a Gaussian process (\GP{}). In the case that multiple properties are measured for a given point in the design space, multiple independent single-property \GP{} models are used. A \GP{} is best conceptualized as a probability distribution over plausible functions. In the absence of data, we can define a prior distribution over functions (Equation \ref{eqn:prior}). The mean of the prior distribution is assigned the value $\mathbf{0}$ everywhere in the domain and the prior covariance function is denoted $\mathbf{K}$ and is derived from the squared exponential kernel (See Methods):
\begin{equation}
    p(f) = \mathcal{GP}(f ; \mathbf{0}, \mathbf{K}).
\label{eqn:prior}
\end{equation}

As data are collected, the prior distribution is updated to the posterior distribution, $p(f | \mathcal{D}_t)$. The mean and marginal standard deviation of this new distribution are termed the posterior mean function (\posteriormean{}) and posterior standard deviation function ($f^{\sigma}$), respectively. An example of the posterior mean function (shown in red) and posterior standard deviation function (shown as blue band) is shown in Figure \ref{fig:approach_overview}B based on a training dataset of five design points and corresponding measured properties. Note, that the \GP{} model estimates low uncertainties near measured points.

It is possible to sample from a \GP{} to yield a series of statistically consistent plausible fitting functions (termed posterior function samples), \posteriorsamples{} (Equation \ref{eqn:posteriorsamples}), and displayed as blue curves in Figure \ref{fig:approach_overview}B. We denote these as
\begin{equation}
    f_i \sim p(f | \mathcal{D}_t)\,.
    \label{eqn:posteriorsamples}
\end{equation}

Given a trained $\mathcal{GP}$ model, an algorithm (such as the Level Band algorithm) can be executed on the posterior mean function or on a posterior function sample (Equation \ref{eqn:executeGP} and Figure \ref{fig:approach_overview}C):
\begin{equation}
    \mathcal{A}(\bar{f}, X) \rightarrow \mathcal{\bar{T}}, \quad
    \mathcal{A}(f_i, X) \rightarrow \mathcal{T}_i.
    \label{eqn:executeGP}
\end{equation}

Note that the algorithm execution step is both fast and inexpensive, as it does not require any additional measurements to be performed. The subsets returned by the algorithm are two different types of \textit{predictions} of the identity of the ground-truth (and unknown) target subset $\mathcal{T}_*$.

The information obtained from the execution of the algorithms can be used to build a guiding function, termed the acquisition function. Each point in the design space is assigned an acquisition value quantifying the relative importance for subsequent measurement. The next design point to measure is the one with the highest acquisition value (Figure \ref{fig:approach_overview}D). Specific acquisition functions for the task of BAX are described in the next section. Overall, the data acquisition pipeline follows these steps: 

\begin{enumerate}
    \item Construct an algorithm, $\mathcal{A}(f, X)$, corresponding to a stated experimental goal.
    \item Approximate $f_*$ using a set of independent $\mathcal{GP}$ models trained on limited measurements.
    \item Execute the algorithm on either the $\mathcal{GP}$ posterior mean, \posteriormean, or its posterior samples, \posteriorsamples{}, over the full design space (Equation \ref{eqn:executeGP}). This yields a set of design points that are predicted to conform to the experimental goal. 
    \item Use the algorithm outputs to build a goal-aware acquisition function. 
    \item Perform an experiment on the design point with the highest acquisition function and repeat from step 2.
\end{enumerate}

\subsection*{Multi-property BAX Acquisition Functions}

We present three acquisition functions for the task of BAX: MeanBAX (based on the posterior mean function), InfoBAX (based on the posterior function samples) and SwitchBAX, which dynamically combines the two. In MeanBAX, the user-algorithm, \alg{}, is executed on the posterior mean of the \GP{} model. Here, the output of the algorithm corresponds to the set of points in the design space that are \textit{predicted} to satisfy the experimental goal. For MeanBAX, the acquisition function is equal to the average (across the different measured properties) output marginal standard deviation of the \GP{} models for points in the design space that are \textit{predicted} to be part of the target subset (Equation \ref{eqn:MeanBAX}). The acquisition function is zero for all other points in the design space (Figure \ref{fig:approach_overview}D, bottom panel). Similar single-property variants of this acquisition function have been proposed in other works for specialized applications \cite{torres2019low,katsube2020experimental,tian2021determining,dai2020efficient}. For the MeanBAX algorithm as presented above, two situations often occur that lead to pathological sampling behavior. The first is when no design points are predicted to be in the target subset (i.e. when $\bar{T}^x = \emptyset$); under this condition, the acquisition function is zero across the entire domain. In the second case, the predicted target set may have already been collected (i.e. when $\bar{T}^x \subseteq \mathcal{D}^x_t$); under this condition, the algorithm is forced to repeat queries. Therefore, if either condition is met, we use a default strategy of $\frac{1}{m}\sum^{m}_{j=1} f^\sigma(x)_j$ across the \textit{entire} domain (i.e., Uncertainty Sampling). We therefore define
\begin{equation}
\text{MeanBAX}(x) =
\begin{cases}
     \frac{1}{m}\sum^{m}_{j=1} f^\sigma(x)_j & \text{if } x \in \mathcal{\bar{T}}^x \\
    0 & \text{else } 
\end{cases}
\label{eqn:MeanBAX}
\end{equation}

For the InfoBAX strategy,  the user-algorithm is executed on a series of \GP{} posterior function samples, each yielding a different set of predicted target points (Figure \ref{fig:approach_overview}D, top panel). Since the algorithm output for each posterior function sample may yield a different number of design points, this is not a trivial extension of MeanBAX and requires combining the outputs in a statistically reasonable manner. The InfoBAX acquisition function is defined as
\begin{equation}
\text{InfoBAX}(x) = \frac{1}{m} \sum_j^{m} \left( \mathbf{H}\left[p(y_j | x, \mathcal{D}_t)\right] - \frac{1}{n} \sum_i^n \mathbf{H} \left[ p(y_j | x, \mathcal{D}_t \cup \mathcal{T}_i) \right] \right)
\label{eqn:infoBAX}
\end{equation}

The first term in the InfoBAX acquisition function (Equation~\ref{eqn:infoBAX}) is the entropy (spread) of the posterior predictive distribution. The posterior predictive distribution, $p(y | \mathcal{D}_t)$, is closely related to the posterior distribution, $p(f | \mathcal{D}_t)$, and includes the effect of measurement noise: $p(y | \mathcal{D}_t) = \int p(f | \mathcal{D}_t) p(y | f) dy$. This term essentially performs Uncertainty Sampling and aims to suggest the design point with highest predicted average uncertainty. 

The second term captures the experimental goal through the output of a user-algorithm. For each of the $n$ posterior function samples, a new \GP{} model is trained with the measured dataset plus a predicted dataset corresponding to algorithm execution (i.e. $\mathcal{D}_t \cup \mathcal{T}_i$). Importantly, the predicted datasets and corresponding updated \GP{} models are only used to calculate the acquisition function and are discarded after selecting the next real design point to measure. The entropy of each model is calculated across the design space and then averaged over the number of posterior samples. Finally, to account for the multi-property case, an average is taken over the $m$ properties. Conceptually, InfoBAX relates variance in the \GP{} posterior samples to variance in the algorithm outputs; the acquisition function selects points in the design space where both the models are uncertain AND where that uncertainty influences the algorithm output. For further details, refer to Neiswanger et al. (2021) \cite{neiswanger2021bayesian}. 

Finally, the SwitchBAX strategy is a modification to the MeanBAX strategy which changes the default behavior to InfoBAX rather than US under the condition that either 1) no points are predicted to be in the target subset or 2) all predicted points have already been measured. Based on these conditions, the method dynamically switches between InfoBAX and MeanBAX to guide decision making. 

We refer to the BAX strategies (InfoBAX, MeanBAX, and SwitchBAX) as `goal-aware' because the acquisition function incorporates the user goal directly via algorithm specification. We can compare these approaches to typical acquisition functions for searching a multi-property design space: Random Sampling (RS), Uncertainty Sampling (US), and Expected Hypervolume Improvement (EHVI). RS selects a design point uniformly at random (here, without replacement) from the discrete design space at each iteration. For US, the acquisition function is simply the predicted average standard deviation of the \GP{} models, $\frac{1}{m}\sum^{m}_{j=1} f^\sigma(x)_j$. Intuitively, this corresponds to making measurements where the model is most uncertain about the average value of the measured properties. These two acquisition functions are often used for mapping, in which the goal is to estimate the value of the measured properties over the full design space. EHVI is a specialized multi-objective Bayesian optimization acquisition function that is designed for the specific goal of Pareto front estimation. 

The utility of our multi-property BAX framework is that acquisition functions can be aligned to arbitrarily complex questions about an experimental system. As long as an algorithm that could be executed on the true function can be written, the BAX strategies circumvent the lack of knowledge of the true function by running the algorithm on function samples or the mean of a surrogate model (Figure \ref{fig:approach_overview}). 

\subsection*{Metrics for Sequential Experimental Design}

\begin{figure}[H]
\centering
\includegraphics[width=0.75\linewidth]{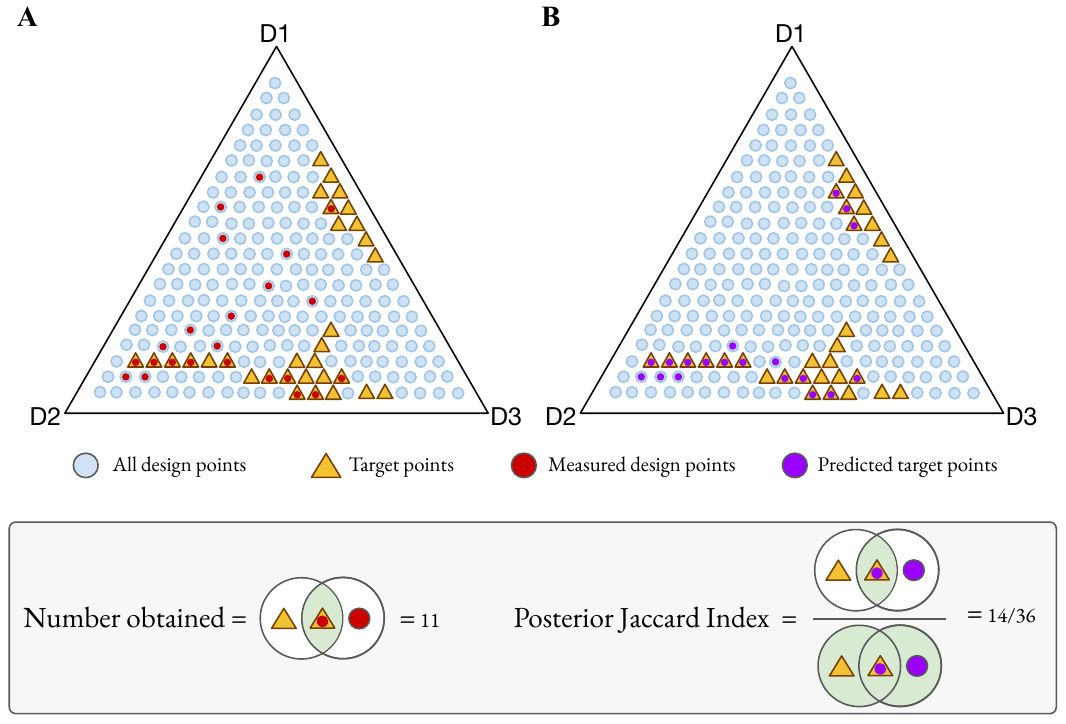}
\caption{Example calculation of the \nobtained{} and \Jposterior{} metrics. The true target subset (gold triangles) of the design space is defined with respect to a specific user goal. (\textbf{A}) Definition of \nobtained{} as the number of measured points (shown in red) that are ground-truth target points. (\textbf{B}) Definition of \Jposterior{} as the intersection over the union of the design points that are predicted to be targets (shown in purple) and the ground truth target points.}
\label{fig:metrics}
\end{figure}

To assess the performance of the various adaptive sampling strategies (RS, US, EHVI, MeanBAX, InfoBAX and SwitchBAX), we introduce two metrics: \nobtained{} and \Jposterior{}. \nobtained{} quantifies the number of measured data points that achieve the experimental goal (Equation \ref{eqn:metrics_nobtained} and Figure \ref{fig:metrics}A), defined as
\begin{equation}
\mathsf{Number} \text{ } \mathsf{Obtained} \text{ } = |\mathcal{D}^x_t \cap \mathcal{T}^x_*|.
\label{eqn:metrics_nobtained}
\end{equation}

The \Jposterior{} quantifies how accurately the \GP{} model knows the ground-truth target subset. To compute this metric, we execute the user-specified algorithm on \posteriormean{} to obtain the set of points that are \textit{predicted} to be in the target subset. This set of points can be compared with the set of points that are actually in the target subset (using the true function). Here, we use the \Jposterior{} (intersection over union), a metric between 0 and 1 which quantifies the degree of set overlap, to compare sets (Equation \ref{eqn:metrics_posterior} and Figure \ref{fig:metrics}B). We define this as
\begin{equation}
\mathsf{Posterior} \text{ } \mathsf{Jaccard} \text{ } \mathsf{Index} = \frac{|\mathcal{T}^x_* \cap \mathcal{\bar{T}}^x|}{|\mathcal{T}^x_* \cup \mathcal{\bar{T}}^x|}.
\label{eqn:metrics_posterior}
\end{equation}

Note that computing the \Jposterior{} assumes that the true target subset of the design space is already known. This information is unknown during a real experiment, and therefore the \Jposterior{} is most useful in a benchmarking context to judge the performance of different acquisition schemes on previously collected data. 

To aid in metric evaluation, we also present upper bounds for the two metrics as a function of the amount of data collected. Under optimal sampling, at each iteration. \nobtained{} increases by one until all the target subset points have been measured. The \Jposterior{} upper bound, in contrast, is 1.0 for each iteration. This would correspond to the exceeding unlikely scenario in which a model initialized with no data perfectly predicts which design points are in the target subset.

\section*{Results}

We use two datasets from the fields of nanoparticle synthesis and magnetic materials to benchmark the performance of various acquisition functions (RS, US, EHVI, InfoBAX, MeanBAX, and SwitchBAX) for the task of targeted subset estimation. The following subsections describe three user-defined algorithms (denoted Library, Multiband and Wishlist Algorithms) relevant to materials application. In Supplementary Section \ref{sec:additional_algorithms} and Figures \SIconditional{}-\SIpercentile{} we present two additional flavors of algorithms: conditional algorithms which safe-guard against unachievable goals, and percentile-based algorithms which avoid needing explicit property thresholds.

\subsection*{Nanoparticle Synthesis}

The nanoparticle synthesis dataset consists of discrete samples (pairs of design points and measured properties) from an empirically fit model of the mapping from synthesis conditions (pH, Temperature, Ti(Teoa)$_2$ concentration, TeoaH$_3$ concentration, Teoa = triethanolamine) to TiO$_2$ nanoparticle size (in nm) and polydispersity ($\%$) \cite{pellegrino2020machine}. We added $1\%$ noise to each measurement to simulate noisy acquisition (See Methods).

\begin{figure*}[t!] 
\centering
\includegraphics[width=\linewidth]{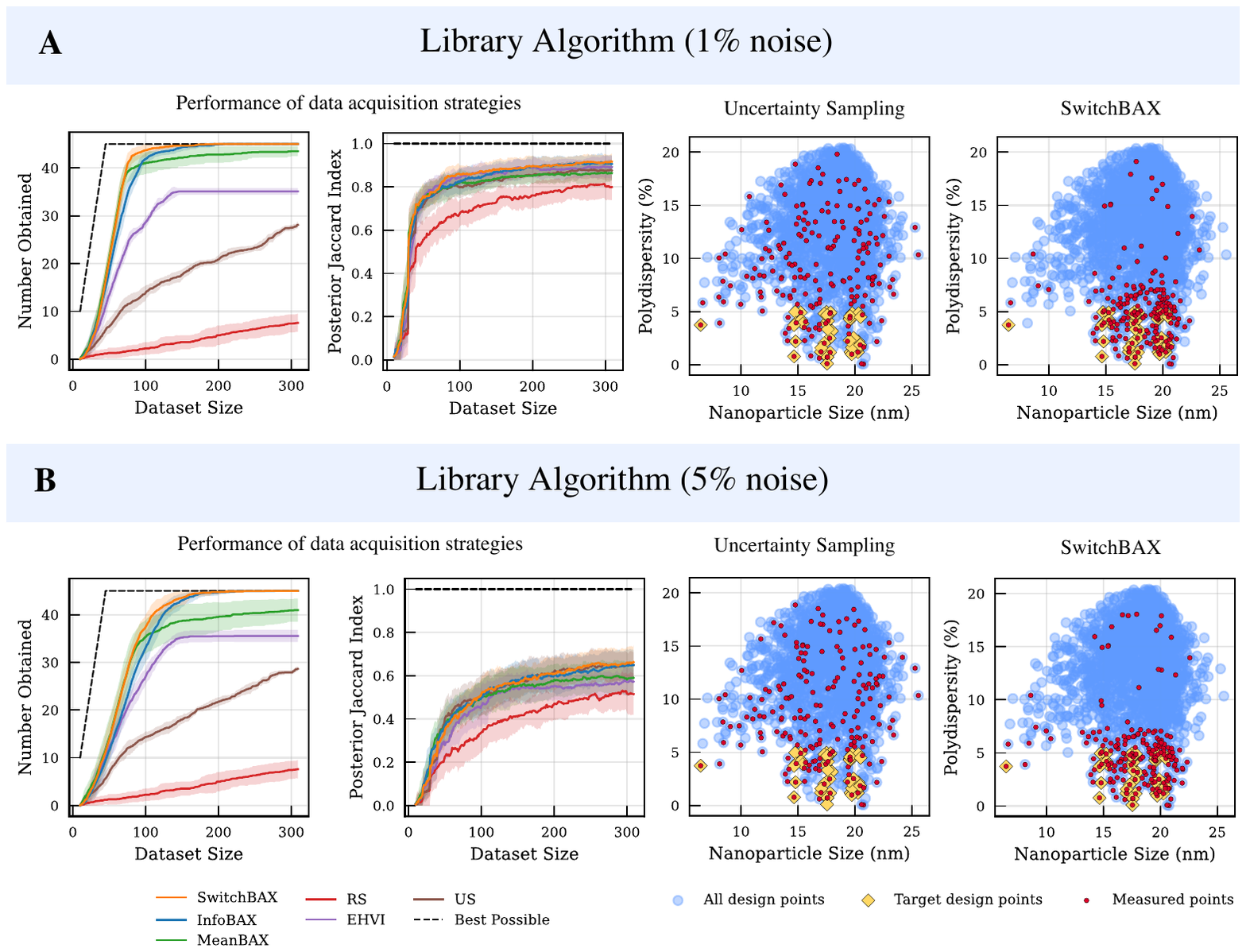}
\caption{Designing an acquisition strategy for the specific goal of finding a library of monodisperse nanoparticles corresponding to the precise specifications: radius $\in$ [6.5, 10,  15, 17.5,  20,  30] $\pm$ 0.5 nm and polydispersity $<$ $5\%$; results are presented for \textbf{A}) 1$\%$ and (\textbf{B}) 5$\%$ noise on the normalized measured properties. BAX strategies, which take into account the user goal, outperform US, RS and EHVI for the  \nobtained{} metric. SwitchBAX samples more densely in the target region relative to US, highlighting the effectiveness of goal-aware sampling.}
\label{fig:np_synthesis}
\end{figure*}

We consider the experimental goal of preparing a library of monodisperse nanoparticles with a series of precisely specified radii; specifically, our aim is to estimate the subset of synthesis conditions that yield nanoparticles with polydispersity $<$ 5$\%$ and where the radius falls into an arbitrarily chosen set of disjoint buckets [6.5, 10,  15, 17.5,  20,  30] $\pm$ 0.5 nm. Such tasks are important as monodisperse nanoparticles of different sizes can be optimal for different catalytic reactions \cite{fong2021utilization,tassone2019aggregation}. It is important to note that this problem is distinct from constrained multi-objective optimization as the goal is to map out \textbf{all} possible syntheses which meet the user specifications. For this example, the user-defined algorithm corresponds to straightforward filtering logic to select the set of disjoint regions which match the stated goal (see Algorithm \ref{algo:library}). 

\begin{algorithm}[H]
\caption{\textsc{Library}: Monodisperse library of nanoparticles.}
\begin{algorithmic}
\Function{Library}{$f$, $X$}
\State $\mathsf{r\_list} = [6.5, 10, 15, 17.5, 20, 30]$ \algorithmiccomment{\textit{Specified target radii}}
\State $\mathsf{r\_tol} = 0.5$ \algorithmiccomment{\textit{Radius tolerance}}
\State $\mathsf{pd\_tol} = 5$ \algorithmiccomment{\textit{Polydispersity threshold}}
\State $\mathcal{T} \gets \emptyset$
\For{$r \in \mathsf{r\_list}$} \algorithmiccomment{\textit{Loop through the set of specified target radii}}
    \State $\mathcal{T}^x_1 \gets \text{argwhere}(\lvert f(X)_1 - r \rvert \leq \mathsf{r\_tol})$ \algorithmiccomment{\textit{Find design points within a given radius band}}
    \State $\mathcal{T}^x_2 \gets \text{argwhere}(f(X)_2 \leq \mathsf{pd\_tol})$ \algorithmiccomment{\textit{Find monodisperse design points}}
    \State $\mathcal{T}^x \gets (\mathcal{T}^x_1 \cap \mathcal{T}^x_2) \cup \mathcal{T}^x$ \algorithmiccomment{\textit{Append monodisperse nanoparticles of fixed radius}}
\EndFor
\State $\mathcal{T} \gets \{\mathcal{T}^x, f(\mathcal{T}^x)\}$
\State \Return $\mathcal{T}$
\EndFunction
\end{algorithmic}
\label{algo:library}
\end{algorithm}

We benchmarked the performance of RS, US, EHVI, MeanBAX, InfoBAX, and SwitchBAX on the library estimation task, using \nobtained{} and \Jposterior{} as metrics (Figure \ref{fig:np_synthesis}A-B). Error bars in these plots correspond to 20 repeats of data acquisition starting with different sets of ten initial points. The BAX strategies outperform RS, US and EHVI in terms of the realistic-setting metric (\nobtained{}) and perform similarly to US in terms of the \Jposterior{} here. 

On average, InfoBAX gives superior long term performance relative to MeanBAX for both metrics. However, MeanBAX performs well initially in terms of the \nobtained{} metric. The SwitchBAX algorithm appears to perform well across both dataset size regimes. The measured properties corresponding to the design points collected under US and InfoBAX are shown in Figure \ref{fig:np_synthesis}. Here, US samples widely in property space and not necessarily in the subset of interest. In contrast, InfoBAX typically samples in regions close to the target subset of points (gold diamonds), showing the effectiveness of user-directed acquisition. Sampling in measured property space for RS, EHVI and MeanBAX are shown in Figure \SIysampling{}.  A t-distributed stochastic neighbor embedding (TSNE) visualization of the sampling in design space is shown in Figure \SIxsampling{} for US and SwitchBAX. 

We also characterized the performance of the acquisition strategies under conditions of higher noise (5$\%$) on the measured properties. Under these conditions, it takes longer to obtain all the target design points for all acquisition strategies. In addition, the \GP{} model is less confident about the location of the target subset of the design space (lower \Jposterior{} relative to 1$\%$). MeanBAX exhibits higher variance in \nobtained{}, while InfoBAX and SwitchBAX appear to be relatively robust to different initializations. Results for additional noise levels (0$\%$ and 10$\%$) are shown in Figure \SInoise{}.

\subsection*{Magnetic Property Estimation}

The magnetic materials characterization dataset consists of a design space of 921 ternary compositions approximately evenly spaced across the ferromagnetic Fe-Co-Ni ternary alloy system \cite{yoo2006identification}. The output measured properties for each composition are the Kerr rotation and the coercivity. The Kerr rotation is a surface-sensitive measure of a material's magnetic properties. Searching for materials with high Kerr rotation is a route to discovering materials for erasable optical recordings \cite{antonov1997computationally}. Coercivity is the field required in a hysteresis loop to completely demagnetize a ferromagnet. The higher the coercivity the less susceptible a particular magnetization state is to flipping due to defects or other mechanisms. 

\begin{figure*}[t!] % this float will be placed at top of p. 2
\centering
\includegraphics[width=\linewidth]{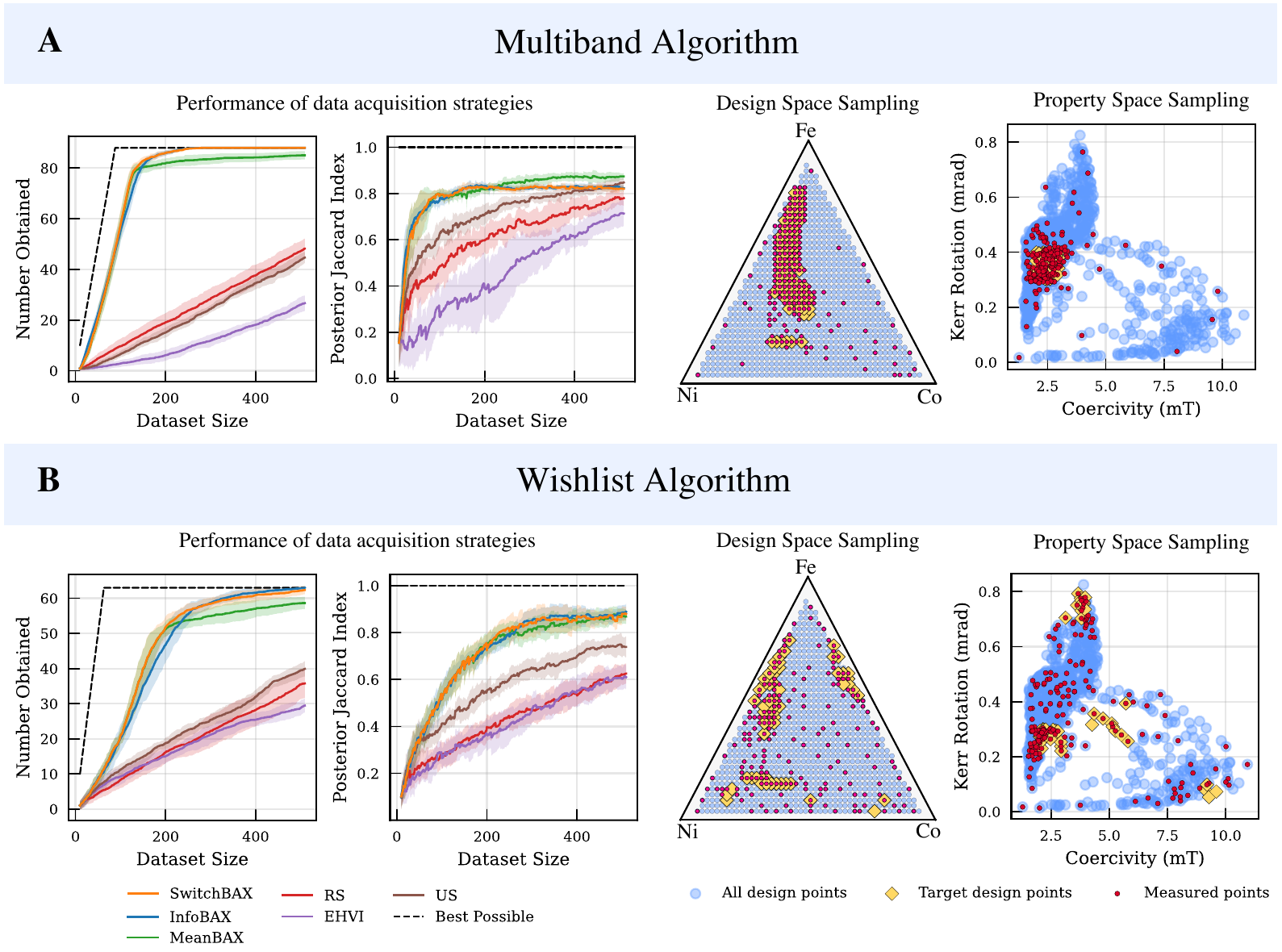}
\caption{Designing an acquisition strategy for the specific goals of finding ternary compositions corresponding to a multiband, [[2.0, 3.0], [0.3, 0.4]] \textbf{A} and to a wishlist, [[2.0,3.0], [0.2, 0.3]] or [[4.0,6.0], [0.2, 0.4]] or [[9.0, 10.0], [0.0, 0.1]] or [[3.0,4.0], [0.7, 0.8]] \textbf{B}; the notation [[a, b], [c, d]] denotes $a$ $<$ Kerr Rotation (mrad) $<$ $b$ and $c$ $<$ coercivity (mT) $<$ $d$. The error bars characterize the robustness to different randomly chosen sets of initial data (one standard deviation computed over 20 repetitions with 10 initial datapoints). Design and property space sampling patterns are shown after 200 iterations for the SwitchBAX acquisition function. The BAX strategies show superior sampling profiles relative to RS, US and EHVI, underscoring the utility of user-directed algorithmic sampling.}
\label{fig:ternary}
\end{figure*}

For this dataset, we highlight two algorithms: the Multiband Algorithm (an intersection of two level bands) and the Wishlist Algorithm (a composition of multiple multibands). 

\subsubsection*{Multiband Algorithm}

The Multiband Algorithm aims to estimate the region of the design space where the measurable properties falls within a separate user-defined band. This goal can be simply expressed by a filtering algorithm which checks for the intersection of the target subsets for each measured property (see Algorithm \ref{algo:levelbandintersection}). Here, the stated experimental goal is to determine the set of design points for which the coercivity falls in the [2.0, 3.0] mT range and the Kerr Rotation falls in the [0.3, 0.4] mrad range; we employ a shorthand $[[a, b], [c, d]] =$[[2.0, 3.0], [0.3, 0.4]] to describe this region. 

Similarly to the nanoparticle synthesis example, goal-driven acquisition functions (InfoBAX and MeanBAX) perform well relative to RS, US, and EHVI (Figure \ref{fig:ternary}A). EHVI exhibits notably poor performance as it targets a disjoint partition of the design space. Again in this example, we see that MeanBAX performs the best in the short term, while InfoBAX has superior long term performance. Here, it is worth noting that although the desired region is tightly clustered in measured property space, it is more disperse in the design space (Figure \ref{fig:ternary}A). 

\begin{algorithm}[H]
\caption{\textsc{Multiband}: Intersection of two level bands}
\begin{algorithmic}
\Function{Multiband}{$f$, $X$} 
    \State $\mathcal{T}^x_1 \gets \text{argwhere}(a \leq f(X)_1 \leq b)$ \algorithmiccomment{\textit{Level band for property 1}}
    \State $\mathcal{T}^x_2 \gets \text{argwhere}(c \leq f(X)_2 \leq d)$  \algorithmiccomment{\textit{Level band for property 2}}
    \State $\mathcal{T}^x \gets \mathcal{T}^x_1 \cap \mathcal{T}^x_2$ \algorithmiccomment{\textit{Intersection of both level bands}}
    \State $\mathcal{T} \gets \{\mathcal{T}^x, f(\mathcal{T}^x) \}$
    \State \Return $\mathcal{T}$
\EndFunction
\end{algorithmic}
\label{algo:levelbandintersection}
\end{algorithm}

\subsubsection*{Wishlist Algorithm}

The Wishlist Algorithm is a composition (or union) of a series of multibands. It addresses the case that a user may have a variety of experimental goals to realize in an experimental system (but not necessarily mapping). In this specific example, we target the following sets of multiband regions: [[2.0,3.0], [0.2, 0.3]] or [[4.0,6.0], [0.2, 0.4]] or [[9.0, 10.0], [0.0, 0.1]] or [[3.0,4.0], [0.7, 0.8]]. Here, it is notable that the ground-truth target subset is disjoint in design space, making this problem significantly more challenging than the multiband scenario. MeanBAX, InfoBAX, and SwitchBAX again perform well relative to RS, US, and EHVI. Note, in particular, the BAX strategies have a much higher \Jposterior{} relative to US (Figure \ref{fig:ternary}B).

\begin{algorithm}[H]
\caption{\textsc{Wishlist}: Composition of a series of multibands}
\begin{algorithmic}
\Function{Wishlist}{$f$, $X$}
    \State $\mathcal{T}^x$ = $\emptyset$
    % \For{$W([a, b, c, d]) \in [W_1, W_2, ..., W_n]$}
    % \For{$W \in [W_1, W_2, ..., W_n]$}
    \For{$\ell \in \{1, \ldots, L\}$} \algorithmiccomment{\textit{Loop through specified multibands}}
    % \State Set $a, b, c, d$ as lower/upper bounds.
    \State $\mathcal{T}^x_\ell \gets \textsc{Multiband}_\ell(f, X)$ \algorithmiccomment{\textit{Target set corresponding to a single multiband}}
    \State $\mathcal{T}^x \gets \mathcal{T}^x_\ell \cup \mathcal{T}^x$ \algorithmiccomment{\textit{Add current multiband to output set}}
    \EndFor
    \State $\mathcal{T} = \{ \mathcal{T}^x, f(\mathcal{T}^x) \}$
    \State \Return $\mathcal{T}$
\EndFunction
\end{algorithmic}
\label{algo:wishlist}
\end{algorithm}

For the given wishlist example, there exists at least one design point which falls into each of the separate multibands. However, in practical experiments, there are scenarios where satisfying the experimental goal is actually unachievable (i.e. there do not exist any design points which satisfy the goal for one or many multibands). We consider this case in more detail in Supplementary Section \ref{sec:additional_algorithms} section and Figure \SIconditional{} where we showcase a more robust type of algorithm (conditional algorithms) that is capable of dynamically switching strategies based on whether the \GP{} models predict whether the goal is achievable. Notably, this non-trivial change in sampling behavior is enabled by only a minimal change to the algorithm. 

\section*{Discussion}

Efficiently exploring a design space to find materials candidates with precisely specified measured properties is of fundamental importance to future materials innovation and discovery. While there are existing approaches for finding certain target subsets of the design space, such as Bayesian optimization for identifying global minima or Uncertainty Sampling for full-function estimation, the general task of subset estimation has not been studied within a materials context. In this study, we present a multi-property version of Bayesian Algorithm Execution (BAX) to develop sequential decision-making strategies aimed at estimating user-specified target subsets of the design space.

Users can encode their target subset using a simple algorithm that requires only a few lines of code. This algorithm is then automatically converted to goal-aware acquisition functions (InfoBAX, MeanBAX, and SwitchBAX) capable of goal-aware exploration. We evaluated BAX strategies on datasets from the fields of nanoparticle synthesis and magnetic materials. For each case, we retrospectively analyzed the performance of different acquisition strategies using metrics that characterize the number of successful experiments (\nobtained{}) and that characterize the quality of the predictive models in the ground-truth target subset (\Jposterior{}). 

For the nanoparticle synthesis example, we target a non-trivial experimental goal: determining synthesis conditions to develop a library of monodisperse nanoparticles. We observed that the BAX strategies significantly outperformed goal-agnostic RS and US (Figure \ref{fig:np_synthesis}A-B) in terms of \nobtained{}. This result highlights that incorporating the experimental goal into an algorithm allows for a more targeted and efficient sequence of experimental measurements. Here, EHVI, an algorithm designed for an alternate goal (Pareto front estimation), seems like a reasonable strategy for the \Jposterior{} and the \nobtained{} metrics. However, EHVI performs worse on \nobtained{} relative to the BAX strategies, mainly because it misses target points with low polydispersity but high nanoparticle size, due to goal misalignment (Figure \SIysampling{}). In this case, both US and EHVI perform well with respect to the \Jposterior{}, indicating that in this example, a trained \GP{} model is able to learn a good overall model of the search space in a small number of queries. For this specific example, the mapping is relatively smooth as it is derived from a model fit on experimental data and therefore a generic \GP{} model has substantial predictive power across the full design space. This is not generally true for more complex datasets.

For the magnetic materials dataset (real experimental measurements), we introduce two tasks: multiband and wishlist estimation. Once again, the BAX acquisition functions perform favorably when compared to RS, US, and EHVI in terms of \nobtained{}. Additionally, the BAX strategies demonstrate superior performance on the \Jposterior{} metric. This suggests that while techniques such as US and RS can effectively reduce uncertainty across the entire design space, they may not target the reduction of uncertainty in specific regions of interest. In contrast, \GP{} models trained on data acquired from BAX sampling strategies are, by construction, more accurate and confident in the specific target subset, forgoing accurate modelling in the rest of the design space. This key result highlights that efficient data collection requires targeted design space sampling. In addition to the Multiband and Wishlist Algorithms, we also compare BAX strategies against a state-of-the-art acquisition function designed for a specific goal. In Supplementary Section \ref{sec:EHVI_comparison} and Figure \SIpareto{}, we compare BAX methods against EHVI for the task of Pareto front optimization. We find that all approaches return similar results for this dataset, highlighting that, even in cases where acquisition functions have been designed for a given task (i.e. EHVI for Pareto front optimization), algorithm-based approaches can perform comparably. 

Although both MeanBAX, InfoBAX, and SwitchBAX are acquisition functions for the task of Bayesian Algorithm Execution, they exhibit qualitatively different behaviors. We generally see that MeanBAX tends to perform better in the short term, but takes longer to fully estimate the target subset of interest (Figures \ref{fig:np_synthesis}-\ref{fig:ternary}). This finding can be rationalized under the exploitation/exploration trade off. MeanBAX is more exploitative by design due to its use of the posterior mean function. In cases where the posterior mean prediction closely models the true function, MeanBAX will acquire a target point at each iteration. However, the long term performance of MeanBAX may not be optimal due to the earlier exploitative queries hindering a detailed understanding of the entire target subset. For experiments which involve low automation and short experimental budgets, such as human-intensive nanomaterials synthesis, a strategy like MeanBAX may be preferable to InfoBAX to quickly obtain solutions that match user specifications.  
Conversely, InfoBAX, derived from posterior function samples, captures the uncertainty in model predictions and is therefore more successful at exploring the entire target subset. In the presence of noisier data or under-fit models, explorative acquisition functions are also expected to be more robust. We observe this phenomenon in the noise analysis of the nanoparticle synthesis dataset in Figure \ref{fig:np_synthesis}B and Figure \SInoise{}. By construction, InfoBAX performs experiments to gain information about the location of the target subset. For this reason, InfoBAX sometimes queries points outside the target subset in order to better understand the overall shape of the target region; practically, this could mean that the \nobtained{} metric suffers at the expense of potentially improving the \Jposterior{}. Applications involving high-throughput synthesis and characterization or facilities with self-driving laboratories \cite{abolhasani2023rise,macleod2020self,macleod2022self,Szymanski2023} may favor the exploratory InfoBAX approach.

We combine the favorable short and long-term performances of MeanBAX and InfoBAX through the dynamic and parameter-free SwitchBAX strategy. Here, the SwitchBAX strategy performs MeanBAX unless there are either (1) no predicted target points or (2) the predicted target points have already been measured. Under either of these scenarios, the strategy switches to InfoBAX. We find that this approach yields the best overall performance for both the small and medium data regimes considered in this work. Interestingly, we also observe a case in Figure \SIfixedhypers{} where initially InfoBAX outperforms MeanBAX. In this scenario, SwitchBAX still performs well; this finding indicates that initial sampling based on InfoBAX can assist later MeanBAX performance. In general, we expect SwitchBAX to outperform MeanBAX since defaulting to InfoBAX is better than defaulting to US; in other words, it is better to explore with a purpose rather than to perform general exploration. However, it is possible that InfoBAX could outperform SwitchBAX for other datasets or user-algorithms, which is an important point to study in future work. 

While BAX strategies generally outperform RS, US, and EHVI for specific target subsets, it may be possible to develop task-specific acquisition functions (like EHVI for Pareto front estimation) that yield equivalent or superior performance. However, creating such acquisition functions requires time and often substantial mathematical insight. Furthermore, these acquisition functions may only be applicable in specific, one-off settings. The power of the BAX framework lies in abstracting custom acquisition function development from the user, making it more accessible for experimentalists to employ specifically targeted search strategies. 

While designed for materials, our method is directly applicable to other fields. We anticipate that our approach will find broad application across the natural and physical sciences in problems involving multidimensional design and property spaces. 

\subsection*{Code Availability}

We provide a user-friendly implementation \cite{chitturi2023} of the three Bayesian algorithm execution strategies at~\url{https://github.com/src47/multibax-sklearn}. This repository contains tutorial notebooks to aid in guiding real experiments. The GPflow code and generated data \cite{chitturi2023gpflow} for this study are also available at~\url{https://github.com/src47/materials-bax-gpflow}.

\section*{Methods}
\subsection*{Modelling}

Independent \GP{} models with zero prior-mean functions and a squared-exponential covariance functions (kernels) were used to model the mapping from the design space to each normalized measured property. The \GP{} modelling was performed using GPflow \cite{GPflow2017}. In addition, Automatic Relevance Determination (ARD) was used, which assigns a different lengthscale, $l_{1:d}$, to each design variable ($X_i$)  (Equation \ref{eqn:kernel}). The exponential kernel encourages the model to predict similar values of the measured property in local regions of the design space. The lengthscale is a hyperparameter which controls the scale of this smooth behavior; a small lengthscale allows the \GP{} to capture large changes for small design space displacements. The kernel variance hyperparameter, $\alpha_{1:m}$, controls the allowable height of each of the $m$ predicted properties from the \GP{} models. The likelihood variance $\sigma_{1:m}$, a hyperparameter which models noise on the measured properties was used and fixed to one of the following constant values (0.0, 0.01, 0.05 or 1.0),
\begin{equation}
k\left(x, x'\right)=\alpha_m \exp \left(-\frac{\left(x -x' \right)^2}{2 \ell_d^2}\right)
\label{eqn:kernel}
\end{equation}
\GP{} hyperparameters for the lengthscales and kernel variances were fit using five-fold cross validation using the log likelihood as the optimization metric. An adaptive hyperparameter fitting scheme was employed in which the hyperparameters were re-fit every ten data points collected.

Each design variable was normalized to the range (0, 1) using min-max scalarization. This normalization is possible as the design space is assumed to be discrete and fully specified and enumerated. Measured properties were normalized to the range (-1, 1). Here, maximum and minimum ranges were estimated based on domain knowledge. For the nanoparticle synthesis dataset, the nanoparticle sizes and polydispersities were assumed to fall in the range of [0, 30] nm and [0, 30] $\%$, respectively. For the magnetic optimization dataset, the measured properties fall between [0,1] mT and [0,10] mrad for the Kerr rotation and coercivity, respectively. 

\subsection*{Datasets}
\subsubsection*{Nanoparticle synthesis}
The nanoparticle synthesis dataset consists of 1997 random settings for the variables $x$ = [$x_1$, $x_2$, $x_3$, $x_4$] (normalized Ti(Teoa)$_2$ concentration, TeoaH$_3$ concentration, pH and T) from an empirically fit model \cite{pellegrino2020machine} for the nanoparticle radius ($y_1$) and the polydispersity ($y_2$) as a function of synthesis parameters:
{\small
\begin{align*}
y_1 &= 19.36549 - 0.2797x_1 + 1.56885x_2 + 3.5447x_3 + 1.82225x_4 \\
&\quad - 1.1978x_1x_2 - 1.66594x_1x_3 - 1.62873x_1x_4 - 0.02003x_2x_3 \\
&\quad - 0.001268x_2x_4 - 0.35086x_3x_4 + 0.3914x_1^2 + 0.52265x_2^2 \\ &\quad  - 0.81701x_3^2 - 2.74921x_4^2
\end{align*}}
{\small
\begin{align*}
y_2 &= 19.6114239+1.0313718x_1+1.48527x_2+1.7991534x_3 \\
&\quad -4.1983899x_4+1.4263262x_1x_2-0.4279443x_1x_3 \\ 
&\quad -1.3865203x_1x_4 -1.051601x_2x_3-2.06380x_2x_4 \\ 
&\quad -2.476674x_3x_4-0.4497319x_1^2 -1.8040123x_2^2 \\
&\quad-3.8699325x_3^2-2.6148x_4^2
\end{align*}
}
Gaussian noise with $\sigma$ = 0.01 or 0.05 was added to the normalized values for $y_1$ and $y_2$ at the point of measurement. Figure \SInoise{} also shows cases with noise levels of 0.0 and 0.1. 

\subsubsection*{Magnetic Property Estimation}
The magnetic materials dataset corresponds to 921 compositions from the Fe-Co-Ni ternary alloy system \cite{yoo2006identification,wang2022benchmarking}. The composition values for each element range from [0, 100]. For each ternary composition, two materials properties are measured: Kerr rotation (mrad) and the coercivity (mT). 

\subsection*{Sequential Design of Experiments}
Data acquisition strategies were compared for the Library, Multiband, and Wishlist Algorithms. We used the Trieste EHVI implementation for multi-objective bayesian optimization \cite{trieste2023}. In general, the following settings were used: 10 random initial datapoints, 20 experimental repeats, adaptive hyperfitting every 10 iterations, and prevention of requerying design points. In Supplementary Section \ref{sec:fixed_hypers} and Figure \SIfixedhypers{} we also show sampling results for \GP{} for fixed hyperparameters. The InfoBAX and SwitchBAX strategies used 15 posterior samples from the \GP{} model for algorithm execution. 300 and 500 datapoints were acquired for the nanoparticle synthesis and magnetic materials datasets, respectively. Visualization of design space sampling use the python-ternary package \cite{pythonternary}. 

% \subsection*{AI Assistance}

% ChatGPT was employed to streamline software development. All AI outputs were thoroughly validated.

\clearpage
\subsection*{Acknowledgements}
This work is supported in part by the U.S. Department of Energy, Office of Science, Office of Basic Energy Sciences under Contract No. DE-AC02-76SF00515. AR and FHJ acknowledge funding from the National Science Foundation (NSF) program Designing Materials to Revolutionize and Engineer our Future (DMREF) via a project DMR-1922312. CJT was supported by the U.S. Department of Energy, Office of Science, Basic Energy Sciences, Chemical Sciences, Geosciences, and Biosciences Division under SLAC Contract No. DE-AC02-76SF00515. The authors thank D. Boe, C. Cheng, S. Gasiorowski, J. Gregoire, T. Lane, W. Michaels, Y. Nashed, M. Robinson, R. Walroth, and C. Wells for manuscript feedback. The authors acknowledge the use of ChatGPT to streamline software development. 

\subsection*{References}
\bibliography{references/all_refs}

%aipnum4-2.bst 2019-01-14 (MD) hand-edited version of apsrev4-1.bst
%Control: key (0)
%Control: author (8) initials jnrlst
%Control: editor formatted (1) identically to author
%Control: production of article title (0) allowed
%Control: page (1) range
%Control: year (1) truncated
%Control: production of eprint (0) enabled
\begin{thebibliography}{67}%
\makeatletter
\providecommand \@ifxundefined [1]{%
 \@ifx{#1\undefined}
}%
\providecommand \@ifnum [1]{%
 \ifnum #1\expandafter \@firstoftwo
 \else \expandafter \@secondoftwo
 \fi
}%
\providecommand \@ifx [1]{%
 \ifx #1\expandafter \@firstoftwo
 \else \expandafter \@secondoftwo
 \fi
}%
\providecommand \natexlab [1]{#1}%
\providecommand \enquote  [1]{``#1''}%
\providecommand \bibnamefont  [1]{#1}%
\providecommand \bibfnamefont [1]{#1}%
\providecommand \citenamefont [1]{#1}%
\providecommand \href@noop [0]{\@secondoftwo}%
\providecommand \href [0]{\begingroup \@sanitize@url \@href}%
\providecommand \@href[1]{\@@startlink{#1}\@@href}%
\providecommand \@@href[1]{\endgroup#1\@@endlink}%
\providecommand \@sanitize@url [0]{\catcode `\\12\catcode `\$12\catcode
  `\&12\catcode `\#12\catcode `\^12\catcode `\_12\catcode `\%12\relax}%
\providecommand \@@startlink[1]{}%
\providecommand \@@endlink[0]{}%
\providecommand \url  [0]{\begingroup\@sanitize@url \@url }%
\providecommand \@url [1]{\endgroup\@href {#1}{\urlprefix }}%
\providecommand \urlprefix  [0]{URL }%
\providecommand \Eprint [0]{\href }%
\providecommand \doibase [0]{https://doi.org/}%
\providecommand \selectlanguage [0]{\@gobble}%
\providecommand \bibinfo  [0]{\@secondoftwo}%
\providecommand \bibfield  [0]{\@secondoftwo}%
\providecommand \translation [1]{[#1]}%
\providecommand \BibitemOpen [0]{}%
\providecommand \bibitemStop [0]{}%
\providecommand \bibitemNoStop [0]{.\EOS\space}%
\providecommand \EOS [0]{\spacefactor3000\relax}%
\providecommand \BibitemShut  [1]{\csname bibitem#1\endcsname}%
\let\auto@bib@innerbib\@empty
%</preamble>
\bibitem [{\citenamefont {Goodenough}\ and\ \citenamefont
  {Park}(2013)}]{goodenough2013li}%
  \BibitemOpen
  \bibfield  {author} {\bibinfo {author} {\bibfnamefont {J.~B.}\ \bibnamefont
  {Goodenough}}\ and\ \bibinfo {author} {\bibfnamefont {K.-S.}\ \bibnamefont
  {Park}},\ }\bibfield  {title} {\enquote {\bibinfo {title} {The li-ion
  rechargeable battery: a perspective},}\ }\href@noop {} {\bibfield  {journal}
  {\bibinfo  {journal} {Journal of the American Chemical Society}\ }\textbf
  {\bibinfo {volume} {135}},\ \bibinfo {pages} {1167--1176} (\bibinfo {year}
  {2013})}\BibitemShut {NoStop}%
\bibitem [{\citenamefont {Lee}, \citenamefont {Nagaosa},\ and\ \citenamefont
  {Wen}(2006)}]{lee2006doping}%
  \BibitemOpen
  \bibfield  {author} {\bibinfo {author} {\bibfnamefont {P.~A.}\ \bibnamefont
  {Lee}}, \bibinfo {author} {\bibfnamefont {N.}~\bibnamefont {Nagaosa}},\ and\
  \bibinfo {author} {\bibfnamefont {X.-G.}\ \bibnamefont {Wen}},\ }\bibfield
  {title} {\enquote {\bibinfo {title} {Doping a mott insulator: Physics of
  high-temperature superconductivity},}\ }\href@noop {} {\bibfield  {journal}
  {\bibinfo  {journal} {Reviews of modern physics}\ }\textbf {\bibinfo {volume}
  {78}},\ \bibinfo {pages} {17} (\bibinfo {year} {2006})}\BibitemShut {NoStop}%
\bibitem [{\citenamefont {Suh}\ \emph {et~al.}(2020)\citenamefont {Suh},
  \citenamefont {Fare}, \citenamefont {Warren},\ and\ \citenamefont
  {Pyzer-Knapp}}]{suh2020evolving}%
  \BibitemOpen
  \bibfield  {author} {\bibinfo {author} {\bibfnamefont {C.}~\bibnamefont
  {Suh}}, \bibinfo {author} {\bibfnamefont {C.}~\bibnamefont {Fare}}, \bibinfo
  {author} {\bibfnamefont {J.~A.}\ \bibnamefont {Warren}},\ and\ \bibinfo
  {author} {\bibfnamefont {E.~O.}\ \bibnamefont {Pyzer-Knapp}},\ }\bibfield
  {title} {\enquote {\bibinfo {title} {Evolving the materials genome: How
  machine learning is fueling the next generation of materials discovery},}\
  }\href@noop {} {\bibfield  {journal} {\bibinfo  {journal} {Annual Review of
  Materials Research}\ }\textbf {\bibinfo {volume} {50}},\ \bibinfo {pages}
  {1--25} (\bibinfo {year} {2020})}\BibitemShut {NoStop}%
\bibitem [{\citenamefont {Montoya}\ \emph {et~al.}(2022)\citenamefont
  {Montoya}, \citenamefont {Aykol}, \citenamefont {Anapolsky}, \citenamefont
  {Gopal}, \citenamefont {Herring}, \citenamefont {Hummelsh{\o}j},
  \citenamefont {Hung}, \citenamefont {Kwon}, \citenamefont {Schweigert},
  \citenamefont {Sun} \emph {et~al.}}]{montoya2022toward}%
  \BibitemOpen
  \bibfield  {author} {\bibinfo {author} {\bibfnamefont {J.~H.}\ \bibnamefont
  {Montoya}}, \bibinfo {author} {\bibfnamefont {M.}~\bibnamefont {Aykol}},
  \bibinfo {author} {\bibfnamefont {A.}~\bibnamefont {Anapolsky}}, \bibinfo
  {author} {\bibfnamefont {C.~B.}\ \bibnamefont {Gopal}}, \bibinfo {author}
  {\bibfnamefont {P.~K.}\ \bibnamefont {Herring}}, \bibinfo {author}
  {\bibfnamefont {J.~S.}\ \bibnamefont {Hummelsh{\o}j}}, \bibinfo {author}
  {\bibfnamefont {L.}~\bibnamefont {Hung}}, \bibinfo {author} {\bibfnamefont
  {H.-K.}\ \bibnamefont {Kwon}}, \bibinfo {author} {\bibfnamefont
  {D.}~\bibnamefont {Schweigert}}, \bibinfo {author} {\bibfnamefont
  {S.}~\bibnamefont {Sun}}, \emph {et~al.},\ }\bibfield  {title} {\enquote
  {\bibinfo {title} {Toward autonomous materials research: Recent progress and
  future challenges},}\ }\href@noop {} {\bibfield  {journal} {\bibinfo
  {journal} {Applied Physics Reviews}\ }\textbf {\bibinfo {volume} {9}}
  (\bibinfo {year} {2022})}\BibitemShut {NoStop}%
\bibitem [{\citenamefont {Shahriari}\ \emph {et~al.}(2015)\citenamefont
  {Shahriari}, \citenamefont {Swersky}, \citenamefont {Wang}, \citenamefont
  {Adams},\ and\ \citenamefont {De~Freitas}}]{shahriari2015taking}%
  \BibitemOpen
  \bibfield  {author} {\bibinfo {author} {\bibfnamefont {B.}~\bibnamefont
  {Shahriari}}, \bibinfo {author} {\bibfnamefont {K.}~\bibnamefont {Swersky}},
  \bibinfo {author} {\bibfnamefont {Z.}~\bibnamefont {Wang}}, \bibinfo {author}
  {\bibfnamefont {R.~P.}\ \bibnamefont {Adams}},\ and\ \bibinfo {author}
  {\bibfnamefont {N.}~\bibnamefont {De~Freitas}},\ }\bibfield  {title}
  {\enquote {\bibinfo {title} {Taking the human out of the loop: A review of
  bayesian optimization},}\ }\href@noop {} {\bibfield  {journal} {\bibinfo
  {journal} {Proceedings of the IEEE}\ }\textbf {\bibinfo {volume} {104}},\
  \bibinfo {pages} {148--175} (\bibinfo {year} {2015})}\BibitemShut {NoStop}%
\bibitem [{\citenamefont {Kusne}\ \emph {et~al.}(2020)\citenamefont {Kusne},
  \citenamefont {Yu}, \citenamefont {Wu}, \citenamefont {Zhang}, \citenamefont
  {Hattrick-Simpers}, \citenamefont {DeCost}, \citenamefont {Sarker},
  \citenamefont {Oses}, \citenamefont {Toher}, \citenamefont {Curtarolo} \emph
  {et~al.}}]{kusne2020fly}%
  \BibitemOpen
  \bibfield  {author} {\bibinfo {author} {\bibfnamefont {A.~G.}\ \bibnamefont
  {Kusne}}, \bibinfo {author} {\bibfnamefont {H.}~\bibnamefont {Yu}}, \bibinfo
  {author} {\bibfnamefont {C.}~\bibnamefont {Wu}}, \bibinfo {author}
  {\bibfnamefont {H.}~\bibnamefont {Zhang}}, \bibinfo {author} {\bibfnamefont
  {J.}~\bibnamefont {Hattrick-Simpers}}, \bibinfo {author} {\bibfnamefont
  {B.}~\bibnamefont {DeCost}}, \bibinfo {author} {\bibfnamefont
  {S.}~\bibnamefont {Sarker}}, \bibinfo {author} {\bibfnamefont
  {C.}~\bibnamefont {Oses}}, \bibinfo {author} {\bibfnamefont {C.}~\bibnamefont
  {Toher}}, \bibinfo {author} {\bibfnamefont {S.}~\bibnamefont {Curtarolo}},
  \emph {et~al.},\ }\bibfield  {title} {\enquote {\bibinfo {title} {On-the-fly
  closed-loop materials discovery via bayesian active learning},}\ }\href@noop
  {} {\bibfield  {journal} {\bibinfo  {journal} {Nature communications}\
  }\textbf {\bibinfo {volume} {11}},\ \bibinfo {pages} {5966} (\bibinfo {year}
  {2020})}\BibitemShut {NoStop}%
\bibitem [{\citenamefont {Kochenderfer}\ and\ \citenamefont
  {Wheeler}(2019)}]{kochenderfer2019algorithms}%
  \BibitemOpen
  \bibfield  {author} {\bibinfo {author} {\bibfnamefont {M.~J.}\ \bibnamefont
  {Kochenderfer}}\ and\ \bibinfo {author} {\bibfnamefont {T.~A.}\ \bibnamefont
  {Wheeler}},\ }\href@noop {} {\emph {\bibinfo {title} {Algorithms for
  optimization}}}\ (\bibinfo  {publisher} {Mit Press},\ \bibinfo {year}
  {2019})\BibitemShut {NoStop}%
\bibitem [{\citenamefont {Dave}\ \emph {et~al.}(2020)\citenamefont {Dave},
  \citenamefont {Mitchell}, \citenamefont {Kandasamy}, \citenamefont {Wang},
  \citenamefont {Burke}, \citenamefont {Paria}, \citenamefont {P{\'o}czos},
  \citenamefont {Whitacre},\ and\ \citenamefont
  {Viswanathan}}]{dave2020autonomous}%
  \BibitemOpen
  \bibfield  {author} {\bibinfo {author} {\bibfnamefont {A.}~\bibnamefont
  {Dave}}, \bibinfo {author} {\bibfnamefont {J.}~\bibnamefont {Mitchell}},
  \bibinfo {author} {\bibfnamefont {K.}~\bibnamefont {Kandasamy}}, \bibinfo
  {author} {\bibfnamefont {H.}~\bibnamefont {Wang}}, \bibinfo {author}
  {\bibfnamefont {S.}~\bibnamefont {Burke}}, \bibinfo {author} {\bibfnamefont
  {B.}~\bibnamefont {Paria}}, \bibinfo {author} {\bibfnamefont
  {B.}~\bibnamefont {P{\'o}czos}}, \bibinfo {author} {\bibfnamefont
  {J.}~\bibnamefont {Whitacre}},\ and\ \bibinfo {author} {\bibfnamefont
  {V.}~\bibnamefont {Viswanathan}},\ }\bibfield  {title} {\enquote {\bibinfo
  {title} {Autonomous discovery of battery electrolytes with robotic
  experimentation and machine learning},}\ }\href@noop {} {\bibfield  {journal}
  {\bibinfo  {journal} {Cell Reports Physical Science}\ }\textbf {\bibinfo
  {volume} {1}} (\bibinfo {year} {2020})}\BibitemShut {NoStop}%
\bibitem [{\citenamefont {Greenhill}\ \emph {et~al.}(2020)\citenamefont
  {Greenhill}, \citenamefont {Rana}, \citenamefont {Gupta}, \citenamefont
  {Vellanki},\ and\ \citenamefont {Venkatesh}}]{greenhill2020bayesian}%
  \BibitemOpen
  \bibfield  {author} {\bibinfo {author} {\bibfnamefont {S.}~\bibnamefont
  {Greenhill}}, \bibinfo {author} {\bibfnamefont {S.}~\bibnamefont {Rana}},
  \bibinfo {author} {\bibfnamefont {S.}~\bibnamefont {Gupta}}, \bibinfo
  {author} {\bibfnamefont {P.}~\bibnamefont {Vellanki}},\ and\ \bibinfo
  {author} {\bibfnamefont {S.}~\bibnamefont {Venkatesh}},\ }\bibfield  {title}
  {\enquote {\bibinfo {title} {Bayesian optimization for adaptive experimental
  design: A review},}\ }\href@noop {} {\bibfield  {journal} {\bibinfo
  {journal} {IEEE access}\ }\textbf {\bibinfo {volume} {8}},\ \bibinfo {pages}
  {13937--13948} (\bibinfo {year} {2020})}\BibitemShut {NoStop}%
\bibitem [{\citenamefont {Emmerich}, \citenamefont {Deutz},\ and\ \citenamefont
  {Klinkenberg}(2011)}]{emmerich2011hypervolume}%
  \BibitemOpen
  \bibfield  {author} {\bibinfo {author} {\bibfnamefont {M.~T.}\ \bibnamefont
  {Emmerich}}, \bibinfo {author} {\bibfnamefont {A.~H.}\ \bibnamefont
  {Deutz}},\ and\ \bibinfo {author} {\bibfnamefont {J.~W.}\ \bibnamefont
  {Klinkenberg}},\ }\bibfield  {title} {\enquote {\bibinfo {title}
  {Hypervolume-based expected improvement: Monotonicity properties and exact
  computation},}\ }in\ \href@noop {} {\emph {\bibinfo {booktitle} {2011 IEEE
  Congress of Evolutionary Computation (CEC)}}}\ (\bibinfo {organization}
  {IEEE},\ \bibinfo {year} {2011})\ pp.\ \bibinfo {pages}
  {2147--2154}\BibitemShut {NoStop}%
\bibitem [{\citenamefont {Daulton}, \citenamefont {Balandat},\ and\
  \citenamefont {Bakshy}(2020)}]{daulton2020differentiable}%
  \BibitemOpen
  \bibfield  {author} {\bibinfo {author} {\bibfnamefont {S.}~\bibnamefont
  {Daulton}}, \bibinfo {author} {\bibfnamefont {M.}~\bibnamefont {Balandat}},\
  and\ \bibinfo {author} {\bibfnamefont {E.}~\bibnamefont {Bakshy}},\
  }\bibfield  {title} {\enquote {\bibinfo {title} {Differentiable expected
  hypervolume improvement for parallel multi-objective bayesian
  optimization},}\ }\href@noop {} {\bibfield  {journal} {\bibinfo  {journal}
  {Advances in Neural Information Processing Systems}\ }\textbf {\bibinfo
  {volume} {33}},\ \bibinfo {pages} {9851--9864} (\bibinfo {year}
  {2020})}\BibitemShut {NoStop}%
\bibitem [{\citenamefont {Daulton}, \citenamefont {Balandat},\ and\
  \citenamefont {Bakshy}(2021)}]{daulton2021parallel}%
  \BibitemOpen
  \bibfield  {author} {\bibinfo {author} {\bibfnamefont {S.}~\bibnamefont
  {Daulton}}, \bibinfo {author} {\bibfnamefont {M.}~\bibnamefont {Balandat}},\
  and\ \bibinfo {author} {\bibfnamefont {E.}~\bibnamefont {Bakshy}},\
  }\bibfield  {title} {\enquote {\bibinfo {title} {Parallel bayesian
  optimization of multiple noisy objectives with expected hypervolume
  improvement},}\ }\href@noop {} {\bibfield  {journal} {\bibinfo  {journal}
  {Advances in Neural Information Processing Systems}\ }\textbf {\bibinfo
  {volume} {34}},\ \bibinfo {pages} {2187--2200} (\bibinfo {year}
  {2021})}\BibitemShut {NoStop}%
\bibitem [{\citenamefont {Knowles}(2006)}]{knowles2006parego}%
  \BibitemOpen
  \bibfield  {author} {\bibinfo {author} {\bibfnamefont {J.}~\bibnamefont
  {Knowles}},\ }\bibfield  {title} {\enquote {\bibinfo {title} {Parego: A
  hybrid algorithm with on-line landscape approximation for expensive
  multiobjective optimization problems},}\ }\href@noop {} {\bibfield  {journal}
  {\bibinfo  {journal} {IEEE transactions on evolutionary computation}\
  }\textbf {\bibinfo {volume} {10}},\ \bibinfo {pages} {50--66} (\bibinfo
  {year} {2006})}\BibitemShut {NoStop}%
\bibitem [{\citenamefont {Wang}\ \emph {et~al.}(2022)\citenamefont {Wang},
  \citenamefont {Liang}, \citenamefont {McDannald}, \citenamefont {Takeuchi},\
  and\ \citenamefont {Kusne}}]{wang2022benchmarking}%
  \BibitemOpen
  \bibfield  {author} {\bibinfo {author} {\bibfnamefont {A.}~\bibnamefont
  {Wang}}, \bibinfo {author} {\bibfnamefont {H.}~\bibnamefont {Liang}},
  \bibinfo {author} {\bibfnamefont {A.}~\bibnamefont {McDannald}}, \bibinfo
  {author} {\bibfnamefont {I.}~\bibnamefont {Takeuchi}},\ and\ \bibinfo
  {author} {\bibfnamefont {A.~G.}\ \bibnamefont {Kusne}},\ }\bibfield  {title}
  {\enquote {\bibinfo {title} {Benchmarking active learning strategies for
  materials optimization and discovery},}\ }\href@noop {} {\bibfield  {journal}
  {\bibinfo  {journal} {Oxford Open Materials Science}\ }\textbf {\bibinfo
  {volume} {2}},\ \bibinfo {pages} {itac006} (\bibinfo {year}
  {2022})}\BibitemShut {NoStop}%
\bibitem [{\citenamefont {Hase}\ \emph {et~al.}(2018)\citenamefont {Hase},
  \citenamefont {Roch}, \citenamefont {Kreisbeck},\ and\ \citenamefont
  {Aspuru-Guzik}}]{hase2018phoenics}%
  \BibitemOpen
  \bibfield  {author} {\bibinfo {author} {\bibfnamefont {F.}~\bibnamefont
  {Hase}}, \bibinfo {author} {\bibfnamefont {L.~M.}\ \bibnamefont {Roch}},
  \bibinfo {author} {\bibfnamefont {C.}~\bibnamefont {Kreisbeck}},\ and\
  \bibinfo {author} {\bibfnamefont {A.}~\bibnamefont {Aspuru-Guzik}},\
  }\bibfield  {title} {\enquote {\bibinfo {title} {Phoenics: a bayesian
  optimizer for chemistry},}\ }\href@noop {} {\bibfield  {journal} {\bibinfo
  {journal} {ACS central science}\ }\textbf {\bibinfo {volume} {4}},\ \bibinfo
  {pages} {1134--1145} (\bibinfo {year} {2018})}\BibitemShut {NoStop}%
\bibitem [{\citenamefont {Rohr}\ \emph {et~al.}(2020)\citenamefont {Rohr},
  \citenamefont {Stein}, \citenamefont {Guevarra}, \citenamefont {Wang},
  \citenamefont {Haber}, \citenamefont {Aykol}, \citenamefont {Suram},\ and\
  \citenamefont {Gregoire}}]{rohr2020benchmarking}%
  \BibitemOpen
  \bibfield  {author} {\bibinfo {author} {\bibfnamefont {B.}~\bibnamefont
  {Rohr}}, \bibinfo {author} {\bibfnamefont {H.~S.}\ \bibnamefont {Stein}},
  \bibinfo {author} {\bibfnamefont {D.}~\bibnamefont {Guevarra}}, \bibinfo
  {author} {\bibfnamefont {Y.}~\bibnamefont {Wang}}, \bibinfo {author}
  {\bibfnamefont {J.~A.}\ \bibnamefont {Haber}}, \bibinfo {author}
  {\bibfnamefont {M.}~\bibnamefont {Aykol}}, \bibinfo {author} {\bibfnamefont
  {S.~K.}\ \bibnamefont {Suram}},\ and\ \bibinfo {author} {\bibfnamefont
  {J.~M.}\ \bibnamefont {Gregoire}},\ }\bibfield  {title} {\enquote {\bibinfo
  {title} {Benchmarking the acceleration of materials discovery by sequential
  learning},}\ }\href@noop {} {\bibfield  {journal} {\bibinfo  {journal}
  {Chemical science}\ }\textbf {\bibinfo {volume} {11}},\ \bibinfo {pages}
  {2696--2706} (\bibinfo {year} {2020})}\BibitemShut {NoStop}%
\bibitem [{\citenamefont {Yamashita}\ \emph {et~al.}(2018)\citenamefont
  {Yamashita}, \citenamefont {Sato}, \citenamefont {Kino}, \citenamefont
  {Miyake}, \citenamefont {Tsuda},\ and\ \citenamefont
  {Oguchi}}]{yamashita2018crystal}%
  \BibitemOpen
  \bibfield  {author} {\bibinfo {author} {\bibfnamefont {T.}~\bibnamefont
  {Yamashita}}, \bibinfo {author} {\bibfnamefont {N.}~\bibnamefont {Sato}},
  \bibinfo {author} {\bibfnamefont {H.}~\bibnamefont {Kino}}, \bibinfo {author}
  {\bibfnamefont {T.}~\bibnamefont {Miyake}}, \bibinfo {author} {\bibfnamefont
  {K.}~\bibnamefont {Tsuda}},\ and\ \bibinfo {author} {\bibfnamefont
  {T.}~\bibnamefont {Oguchi}},\ }\bibfield  {title} {\enquote {\bibinfo {title}
  {Crystal structure prediction accelerated by bayesian optimization},}\
  }\href@noop {} {\bibfield  {journal} {\bibinfo  {journal} {Physical Review
  Materials}\ }\textbf {\bibinfo {volume} {2}},\ \bibinfo {pages} {013803}
  (\bibinfo {year} {2018})}\BibitemShut {NoStop}%
\bibitem [{\citenamefont {Hickman}\ \emph {et~al.}(2022)\citenamefont
  {Hickman}, \citenamefont {Aldeghi}, \citenamefont {H{\"a}se},\ and\
  \citenamefont {Aspuru-Guzik}}]{hickman2022bayesian}%
  \BibitemOpen
  \bibfield  {author} {\bibinfo {author} {\bibfnamefont {R.~J.}\ \bibnamefont
  {Hickman}}, \bibinfo {author} {\bibfnamefont {M.}~\bibnamefont {Aldeghi}},
  \bibinfo {author} {\bibfnamefont {F.}~\bibnamefont {H{\"a}se}},\ and\
  \bibinfo {author} {\bibfnamefont {A.}~\bibnamefont {Aspuru-Guzik}},\
  }\bibfield  {title} {\enquote {\bibinfo {title} {Bayesian optimization with
  known experimental and design constraints for chemistry applications},}\
  }\href@noop {} {\bibfield  {journal} {\bibinfo  {journal} {Digital
  Discovery}\ }\textbf {\bibinfo {volume} {1}},\ \bibinfo {pages} {732--744}
  (\bibinfo {year} {2022})}\BibitemShut {NoStop}%
\bibitem [{\citenamefont {Herbol}\ \emph {et~al.}(2018)\citenamefont {Herbol},
  \citenamefont {Hu}, \citenamefont {Frazier}, \citenamefont {Clancy},\ and\
  \citenamefont {Poloczek}}]{herbol2018efficient}%
  \BibitemOpen
  \bibfield  {author} {\bibinfo {author} {\bibfnamefont {H.~C.}\ \bibnamefont
  {Herbol}}, \bibinfo {author} {\bibfnamefont {W.}~\bibnamefont {Hu}}, \bibinfo
  {author} {\bibfnamefont {P.}~\bibnamefont {Frazier}}, \bibinfo {author}
  {\bibfnamefont {P.}~\bibnamefont {Clancy}},\ and\ \bibinfo {author}
  {\bibfnamefont {M.}~\bibnamefont {Poloczek}},\ }\bibfield  {title} {\enquote
  {\bibinfo {title} {Efficient search of compositional space for hybrid
  organic--inorganic perovskites via bayesian optimization},}\ }\href@noop {}
  {\bibfield  {journal} {\bibinfo  {journal} {npj Computational Materials}\
  }\textbf {\bibinfo {volume} {4}},\ \bibinfo {pages} {51} (\bibinfo {year}
  {2018})}\BibitemShut {NoStop}%
\bibitem [{\citenamefont {Zhang}, \citenamefont {Apley},\ and\ \citenamefont
  {Chen}(2020)}]{zhang2020bayesian}%
  \BibitemOpen
  \bibfield  {author} {\bibinfo {author} {\bibfnamefont {Y.}~\bibnamefont
  {Zhang}}, \bibinfo {author} {\bibfnamefont {D.~W.}\ \bibnamefont {Apley}},\
  and\ \bibinfo {author} {\bibfnamefont {W.}~\bibnamefont {Chen}},\ }\bibfield
  {title} {\enquote {\bibinfo {title} {Bayesian optimization for materials
  design with mixed quantitative and qualitative variables},}\ }\href@noop {}
  {\bibfield  {journal} {\bibinfo  {journal} {Scientific reports}\ }\textbf
  {\bibinfo {volume} {10}},\ \bibinfo {pages} {4924} (\bibinfo {year}
  {2020})}\BibitemShut {NoStop}%
\bibitem [{\citenamefont {Liang}\ \emph {et~al.}(2021)\citenamefont {Liang},
  \citenamefont {Gongora}, \citenamefont {Ren}, \citenamefont {Tiihonen},
  \citenamefont {Liu}, \citenamefont {Sun}, \citenamefont {Deneault},
  \citenamefont {Bash}, \citenamefont {Mekki-Berrada}, \citenamefont {Khan}
  \emph {et~al.}}]{liang2021benchmarking}%
  \BibitemOpen
  \bibfield  {author} {\bibinfo {author} {\bibfnamefont {Q.}~\bibnamefont
  {Liang}}, \bibinfo {author} {\bibfnamefont {A.~E.}\ \bibnamefont {Gongora}},
  \bibinfo {author} {\bibfnamefont {Z.}~\bibnamefont {Ren}}, \bibinfo {author}
  {\bibfnamefont {A.}~\bibnamefont {Tiihonen}}, \bibinfo {author}
  {\bibfnamefont {Z.}~\bibnamefont {Liu}}, \bibinfo {author} {\bibfnamefont
  {S.}~\bibnamefont {Sun}}, \bibinfo {author} {\bibfnamefont {J.~R.}\
  \bibnamefont {Deneault}}, \bibinfo {author} {\bibfnamefont {D.}~\bibnamefont
  {Bash}}, \bibinfo {author} {\bibfnamefont {F.}~\bibnamefont {Mekki-Berrada}},
  \bibinfo {author} {\bibfnamefont {S.~A.}\ \bibnamefont {Khan}}, \emph
  {et~al.},\ }\bibfield  {title} {\enquote {\bibinfo {title} {Benchmarking the
  performance of bayesian optimization across multiple experimental materials
  science domains},}\ }\href@noop {} {\bibfield  {journal} {\bibinfo  {journal}
  {npj Computational Materials}\ }\textbf {\bibinfo {volume} {7}},\ \bibinfo
  {pages} {188} (\bibinfo {year} {2021})}\BibitemShut {NoStop}%
\bibitem [{\citenamefont {Shields}\ \emph {et~al.}(2021)\citenamefont
  {Shields}, \citenamefont {Stevens}, \citenamefont {Li}, \citenamefont
  {Parasram}, \citenamefont {Damani}, \citenamefont {Alvarado}, \citenamefont
  {Janey}, \citenamefont {Adams},\ and\ \citenamefont
  {Doyle}}]{shields2021bayesian}%
  \BibitemOpen
  \bibfield  {author} {\bibinfo {author} {\bibfnamefont {B.~J.}\ \bibnamefont
  {Shields}}, \bibinfo {author} {\bibfnamefont {J.}~\bibnamefont {Stevens}},
  \bibinfo {author} {\bibfnamefont {J.}~\bibnamefont {Li}}, \bibinfo {author}
  {\bibfnamefont {M.}~\bibnamefont {Parasram}}, \bibinfo {author}
  {\bibfnamefont {F.}~\bibnamefont {Damani}}, \bibinfo {author} {\bibfnamefont
  {J.~I.~M.}\ \bibnamefont {Alvarado}}, \bibinfo {author} {\bibfnamefont
  {J.~M.}\ \bibnamefont {Janey}}, \bibinfo {author} {\bibfnamefont {R.~P.}\
  \bibnamefont {Adams}},\ and\ \bibinfo {author} {\bibfnamefont {A.~G.}\
  \bibnamefont {Doyle}},\ }\bibfield  {title} {\enquote {\bibinfo {title}
  {Bayesian reaction optimization as a tool for chemical synthesis},}\
  }\href@noop {} {\bibfield  {journal} {\bibinfo  {journal} {Nature}\ }\textbf
  {\bibinfo {volume} {590}},\ \bibinfo {pages} {89--96} (\bibinfo {year}
  {2021})}\BibitemShut {NoStop}%
\bibitem [{\citenamefont {H{\"a}se}\ \emph {et~al.}(2021)\citenamefont
  {H{\"a}se}, \citenamefont {Aldeghi}, \citenamefont {Hickman}, \citenamefont
  {Roch},\ and\ \citenamefont {Aspuru-Guzik}}]{hase2021gryffin}%
  \BibitemOpen
  \bibfield  {author} {\bibinfo {author} {\bibfnamefont {F.}~\bibnamefont
  {H{\"a}se}}, \bibinfo {author} {\bibfnamefont {M.}~\bibnamefont {Aldeghi}},
  \bibinfo {author} {\bibfnamefont {R.~J.}\ \bibnamefont {Hickman}}, \bibinfo
  {author} {\bibfnamefont {L.~M.}\ \bibnamefont {Roch}},\ and\ \bibinfo
  {author} {\bibfnamefont {A.}~\bibnamefont {Aspuru-Guzik}},\ }\bibfield
  {title} {\enquote {\bibinfo {title} {Gryffin: An algorithm for bayesian
  optimization of categorical variables informed by expert knowledge},}\
  }\href@noop {} {\bibfield  {journal} {\bibinfo  {journal} {Applied Physics
  Reviews}\ }\textbf {\bibinfo {volume} {8}} (\bibinfo {year}
  {2021})}\BibitemShut {NoStop}%
\bibitem [{\citenamefont {Hanaoka}(2021)}]{hanaoka2021bayesian}%
  \BibitemOpen
  \bibfield  {author} {\bibinfo {author} {\bibfnamefont {K.}~\bibnamefont
  {Hanaoka}},\ }\bibfield  {title} {\enquote {\bibinfo {title} {Bayesian
  optimization for goal-oriented multi-objective inverse material design},}\
  }\href@noop {} {\bibfield  {journal} {\bibinfo  {journal} {Iscience}\
  }\textbf {\bibinfo {volume} {24}} (\bibinfo {year} {2021})}\BibitemShut
  {NoStop}%
\bibitem [{\citenamefont {Karasuyama}\ \emph {et~al.}(2020)\citenamefont
  {Karasuyama}, \citenamefont {Kasugai}, \citenamefont {Tamura},\ and\
  \citenamefont {Shitara}}]{karasuyama2020computational}%
  \BibitemOpen
  \bibfield  {author} {\bibinfo {author} {\bibfnamefont {M.}~\bibnamefont
  {Karasuyama}}, \bibinfo {author} {\bibfnamefont {H.}~\bibnamefont {Kasugai}},
  \bibinfo {author} {\bibfnamefont {T.}~\bibnamefont {Tamura}},\ and\ \bibinfo
  {author} {\bibfnamefont {K.}~\bibnamefont {Shitara}},\ }\bibfield  {title}
  {\enquote {\bibinfo {title} {Computational design of stable and highly
  ion-conductive materials using multi-objective bayesian optimization: Case
  studies on diffusion of oxygen and lithium},}\ }\href@noop {} {\bibfield
  {journal} {\bibinfo  {journal} {Computational Materials Science}\ }\textbf
  {\bibinfo {volume} {184}},\ \bibinfo {pages} {109927} (\bibinfo {year}
  {2020})}\BibitemShut {NoStop}%
\bibitem [{\citenamefont {Hu}\ \emph {et~al.}(2023)\citenamefont {Hu},
  \citenamefont {Wang}, \citenamefont {Du}, \citenamefont {Zou}, \citenamefont
  {Wu}, \citenamefont {Tang}, \citenamefont {Ai}, \citenamefont {Zhou},
  \citenamefont {Chen},\ and\ \citenamefont {Shan}}]{hu2023multi}%
  \BibitemOpen
  \bibfield  {author} {\bibinfo {author} {\bibfnamefont {B.}~\bibnamefont
  {Hu}}, \bibinfo {author} {\bibfnamefont {Z.}~\bibnamefont {Wang}}, \bibinfo
  {author} {\bibfnamefont {C.}~\bibnamefont {Du}}, \bibinfo {author}
  {\bibfnamefont {W.}~\bibnamefont {Zou}}, \bibinfo {author} {\bibfnamefont
  {W.}~\bibnamefont {Wu}}, \bibinfo {author} {\bibfnamefont {J.}~\bibnamefont
  {Tang}}, \bibinfo {author} {\bibfnamefont {J.}~\bibnamefont {Ai}}, \bibinfo
  {author} {\bibfnamefont {H.}~\bibnamefont {Zhou}}, \bibinfo {author}
  {\bibfnamefont {R.}~\bibnamefont {Chen}},\ and\ \bibinfo {author}
  {\bibfnamefont {B.}~\bibnamefont {Shan}},\ }\bibfield  {title} {\enquote
  {\bibinfo {title} {Multi-objective bayesian optimization accelerated design
  of tpms structures},}\ }\href@noop {} {\bibfield  {journal} {\bibinfo
  {journal} {International Journal of Mechanical Sciences}\ }\textbf {\bibinfo
  {volume} {244}},\ \bibinfo {pages} {108085} (\bibinfo {year}
  {2023})}\BibitemShut {NoStop}%
\bibitem [{\citenamefont {Khatamsaz}\ \emph {et~al.}(2023)\citenamefont
  {Khatamsaz}, \citenamefont {Vela}, \citenamefont {Singh}, \citenamefont
  {Johnson}, \citenamefont {Allaire},\ and\ \citenamefont
  {Arr{\'o}yave}}]{khatamsaz2023bayesian}%
  \BibitemOpen
  \bibfield  {author} {\bibinfo {author} {\bibfnamefont {D.}~\bibnamefont
  {Khatamsaz}}, \bibinfo {author} {\bibfnamefont {B.}~\bibnamefont {Vela}},
  \bibinfo {author} {\bibfnamefont {P.}~\bibnamefont {Singh}}, \bibinfo
  {author} {\bibfnamefont {D.~D.}\ \bibnamefont {Johnson}}, \bibinfo {author}
  {\bibfnamefont {D.}~\bibnamefont {Allaire}},\ and\ \bibinfo {author}
  {\bibfnamefont {R.}~\bibnamefont {Arr{\'o}yave}},\ }\bibfield  {title}
  {\enquote {\bibinfo {title} {Bayesian optimization with active learning of
  design constraints using an entropy-based approach},}\ }\href@noop {}
  {\bibfield  {journal} {\bibinfo  {journal} {npj Computational Materials}\
  }\textbf {\bibinfo {volume} {9}},\ \bibinfo {pages} {49} (\bibinfo {year}
  {2023})}\BibitemShut {NoStop}%
\bibitem [{\citenamefont {Xu}\ \emph {et~al.}(2023)\citenamefont {Xu},
  \citenamefont {Liu}, \citenamefont {Piper},\ and\ \citenamefont
  {Hsu}}]{xu2023bayesian}%
  \BibitemOpen
  \bibfield  {author} {\bibinfo {author} {\bibfnamefont {W.}~\bibnamefont
  {Xu}}, \bibinfo {author} {\bibfnamefont {Z.}~\bibnamefont {Liu}}, \bibinfo
  {author} {\bibfnamefont {R.~T.}\ \bibnamefont {Piper}},\ and\ \bibinfo
  {author} {\bibfnamefont {J.~W.}\ \bibnamefont {Hsu}},\ }\bibfield  {title}
  {\enquote {\bibinfo {title} {Bayesian optimization of photonic curing process
  for flexible perovskite photovoltaic devices},}\ }\href@noop {} {\bibfield
  {journal} {\bibinfo  {journal} {Solar Energy Materials and Solar Cells}\
  }\textbf {\bibinfo {volume} {249}},\ \bibinfo {pages} {112055} (\bibinfo
  {year} {2023})}\BibitemShut {NoStop}%
\bibitem [{\citenamefont {Wang}\ \emph {et~al.}(2023)\citenamefont {Wang},
  \citenamefont {Huang}, \citenamefont {Xie}, \citenamefont {Liu},
  \citenamefont {Huo}, \citenamefont {Lin}, \citenamefont {Xin},\ and\
  \citenamefont {Tong}}]{wang2023bayesian}%
  \BibitemOpen
  \bibfield  {author} {\bibinfo {author} {\bibfnamefont {X.}~\bibnamefont
  {Wang}}, \bibinfo {author} {\bibfnamefont {Y.}~\bibnamefont {Huang}},
  \bibinfo {author} {\bibfnamefont {X.}~\bibnamefont {Xie}}, \bibinfo {author}
  {\bibfnamefont {Y.}~\bibnamefont {Liu}}, \bibinfo {author} {\bibfnamefont
  {Z.}~\bibnamefont {Huo}}, \bibinfo {author} {\bibfnamefont {M.}~\bibnamefont
  {Lin}}, \bibinfo {author} {\bibfnamefont {H.}~\bibnamefont {Xin}},\ and\
  \bibinfo {author} {\bibfnamefont {R.}~\bibnamefont {Tong}},\ }\bibfield
  {title} {\enquote {\bibinfo {title} {Bayesian-optimization-assisted discovery
  of stereoselective aluminum complexes for ring-opening polymerization of
  racemic lactide},}\ }\href@noop {} {\bibfield  {journal} {\bibinfo  {journal}
  {Nature Communications}\ }\textbf {\bibinfo {volume} {14}},\ \bibinfo {pages}
  {3647} (\bibinfo {year} {2023})}\BibitemShut {NoStop}%
\bibitem [{\citenamefont {Packwood}\ \emph {et~al.}(2017)\citenamefont
  {Packwood} \emph {et~al.}}]{packwood2017bayesian}%
  \BibitemOpen
  \bibfield  {author} {\bibinfo {author} {\bibfnamefont {D.}~\bibnamefont
  {Packwood}} \emph {et~al.},\ }\href@noop {} {\emph {\bibinfo {title}
  {Bayesian optimization for materials science}}}\ (\bibinfo  {publisher}
  {Springer},\ \bibinfo {year} {2017})\BibitemShut {NoStop}%
\bibitem [{\citenamefont {Yager}\ \emph {et~al.}(2023)\citenamefont {Yager},
  \citenamefont {Majewski}, \citenamefont {Noack},\ and\ \citenamefont
  {Fukuto}}]{yager2023autonomous}%
  \BibitemOpen
  \bibfield  {author} {\bibinfo {author} {\bibfnamefont {K.~G.}\ \bibnamefont
  {Yager}}, \bibinfo {author} {\bibfnamefont {P.~W.}\ \bibnamefont {Majewski}},
  \bibinfo {author} {\bibfnamefont {M.~M.}\ \bibnamefont {Noack}},\ and\
  \bibinfo {author} {\bibfnamefont {M.}~\bibnamefont {Fukuto}},\ }\bibfield
  {title} {\enquote {\bibinfo {title} {Autonomous x-ray scattering},}\
  }\href@noop {} {\bibfield  {journal} {\bibinfo  {journal} {Nanotechnology}\
  }\textbf {\bibinfo {volume} {34}},\ \bibinfo {pages} {322001} (\bibinfo
  {year} {2023})}\BibitemShut {NoStop}%
\bibitem [{\citenamefont {Noack}\ \emph {et~al.}(2019)\citenamefont {Noack},
  \citenamefont {Yager}, \citenamefont {Fukuto}, \citenamefont {Doerk},
  \citenamefont {Li},\ and\ \citenamefont {Sethian}}]{noack2019kriging}%
  \BibitemOpen
  \bibfield  {author} {\bibinfo {author} {\bibfnamefont {M.~M.}\ \bibnamefont
  {Noack}}, \bibinfo {author} {\bibfnamefont {K.~G.}\ \bibnamefont {Yager}},
  \bibinfo {author} {\bibfnamefont {M.}~\bibnamefont {Fukuto}}, \bibinfo
  {author} {\bibfnamefont {G.~S.}\ \bibnamefont {Doerk}}, \bibinfo {author}
  {\bibfnamefont {R.}~\bibnamefont {Li}},\ and\ \bibinfo {author}
  {\bibfnamefont {J.~A.}\ \bibnamefont {Sethian}},\ }\bibfield  {title}
  {\enquote {\bibinfo {title} {A kriging-based approach to autonomous
  experimentation with applications to x-ray scattering},}\ }\href@noop {}
  {\bibfield  {journal} {\bibinfo  {journal} {Scientific reports}\ }\textbf
  {\bibinfo {volume} {9}},\ \bibinfo {pages} {11809} (\bibinfo {year}
  {2019})}\BibitemShut {NoStop}%
\bibitem [{\citenamefont {Szymanski}\ \emph
  {et~al.}(2023{\natexlab{a}})\citenamefont {Szymanski}, \citenamefont
  {Bartel}, \citenamefont {Zeng}, \citenamefont {Diallo}, \citenamefont {Kim},\
  and\ \citenamefont {Ceder}}]{szymanski2023adaptively}%
  \BibitemOpen
  \bibfield  {author} {\bibinfo {author} {\bibfnamefont {N.~J.}\ \bibnamefont
  {Szymanski}}, \bibinfo {author} {\bibfnamefont {C.~J.}\ \bibnamefont
  {Bartel}}, \bibinfo {author} {\bibfnamefont {Y.}~\bibnamefont {Zeng}},
  \bibinfo {author} {\bibfnamefont {M.}~\bibnamefont {Diallo}}, \bibinfo
  {author} {\bibfnamefont {H.}~\bibnamefont {Kim}},\ and\ \bibinfo {author}
  {\bibfnamefont {G.}~\bibnamefont {Ceder}},\ }\bibfield  {title} {\enquote
  {\bibinfo {title} {Adaptively driven x-ray diffraction guided by machine
  learning for autonomous phase identification},}\ }\href@noop {} {\bibfield
  {journal} {\bibinfo  {journal} {npj Computational Materials}\ }\textbf
  {\bibinfo {volume} {9}},\ \bibinfo {pages} {31} (\bibinfo {year}
  {2023}{\natexlab{a}})}\BibitemShut {NoStop}%
\bibitem [{\citenamefont {Kalinin}\ \emph {et~al.}(2021)\citenamefont
  {Kalinin}, \citenamefont {Ziatdinov}, \citenamefont {Hinkle}, \citenamefont
  {Jesse}, \citenamefont {Ghosh}, \citenamefont {Kelley}, \citenamefont
  {Lupini}, \citenamefont {Sumpter},\ and\ \citenamefont
  {Vasudevan}}]{kalinin2021automated}%
  \BibitemOpen
  \bibfield  {author} {\bibinfo {author} {\bibfnamefont {S.~V.}\ \bibnamefont
  {Kalinin}}, \bibinfo {author} {\bibfnamefont {M.}~\bibnamefont {Ziatdinov}},
  \bibinfo {author} {\bibfnamefont {J.}~\bibnamefont {Hinkle}}, \bibinfo
  {author} {\bibfnamefont {S.}~\bibnamefont {Jesse}}, \bibinfo {author}
  {\bibfnamefont {A.}~\bibnamefont {Ghosh}}, \bibinfo {author} {\bibfnamefont
  {K.~P.}\ \bibnamefont {Kelley}}, \bibinfo {author} {\bibfnamefont {A.~R.}\
  \bibnamefont {Lupini}}, \bibinfo {author} {\bibfnamefont {B.~G.}\
  \bibnamefont {Sumpter}},\ and\ \bibinfo {author} {\bibfnamefont {R.~K.}\
  \bibnamefont {Vasudevan}},\ }\bibfield  {title} {\enquote {\bibinfo {title}
  {Automated and autonomous experiments in electron and scanning probe
  microscopy},}\ }\href@noop {} {\bibfield  {journal} {\bibinfo  {journal} {ACS
  nano}\ }\textbf {\bibinfo {volume} {15}},\ \bibinfo {pages} {12604--12627}
  (\bibinfo {year} {2021})}\BibitemShut {NoStop}%
\bibitem [{\citenamefont {Ament}\ \emph {et~al.}(2021)\citenamefont {Ament},
  \citenamefont {Amsler}, \citenamefont {Sutherland}, \citenamefont {Chang},
  \citenamefont {Guevarra}, \citenamefont {Connolly}, \citenamefont {Gregoire},
  \citenamefont {Thompson}, \citenamefont {Gomes},\ and\ \citenamefont {van
  Dover}}]{ament2021autonomous}%
  \BibitemOpen
  \bibfield  {author} {\bibinfo {author} {\bibfnamefont {S.}~\bibnamefont
  {Ament}}, \bibinfo {author} {\bibfnamefont {M.}~\bibnamefont {Amsler}},
  \bibinfo {author} {\bibfnamefont {D.~R.}\ \bibnamefont {Sutherland}},
  \bibinfo {author} {\bibfnamefont {M.-C.}\ \bibnamefont {Chang}}, \bibinfo
  {author} {\bibfnamefont {D.}~\bibnamefont {Guevarra}}, \bibinfo {author}
  {\bibfnamefont {A.~B.}\ \bibnamefont {Connolly}}, \bibinfo {author}
  {\bibfnamefont {J.~M.}\ \bibnamefont {Gregoire}}, \bibinfo {author}
  {\bibfnamefont {M.~O.}\ \bibnamefont {Thompson}}, \bibinfo {author}
  {\bibfnamefont {C.~P.}\ \bibnamefont {Gomes}},\ and\ \bibinfo {author}
  {\bibfnamefont {R.~B.}\ \bibnamefont {van Dover}},\ }\bibfield  {title}
  {\enquote {\bibinfo {title} {Autonomous materials synthesis via hierarchical
  active learning of nonequilibrium phase diagrams},}\ }\href@noop {}
  {\bibfield  {journal} {\bibinfo  {journal} {Science Advances}\ }\textbf
  {\bibinfo {volume} {7}},\ \bibinfo {pages} {eabg4930} (\bibinfo {year}
  {2021})}\BibitemShut {NoStop}%
\bibitem [{\citenamefont {Bogunovic}\ \emph {et~al.}(2016)\citenamefont
  {Bogunovic}, \citenamefont {Scarlett}, \citenamefont {Krause},\ and\
  \citenamefont {Cevher}}]{bogunovic2016truncated}%
  \BibitemOpen
  \bibfield  {author} {\bibinfo {author} {\bibfnamefont {I.}~\bibnamefont
  {Bogunovic}}, \bibinfo {author} {\bibfnamefont {J.}~\bibnamefont {Scarlett}},
  \bibinfo {author} {\bibfnamefont {A.}~\bibnamefont {Krause}},\ and\ \bibinfo
  {author} {\bibfnamefont {V.}~\bibnamefont {Cevher}},\ }\bibfield  {title}
  {\enquote {\bibinfo {title} {Truncated variance reduction: A unified approach
  to bayesian optimization and level-set estimation},}\ }\href@noop {}
  {\bibfield  {journal} {\bibinfo  {journal} {Advances in neural information
  processing systems}\ }\textbf {\bibinfo {volume} {29}} (\bibinfo {year}
  {2016})}\BibitemShut {NoStop}%
\bibitem [{\citenamefont {Ha}\ \emph {et~al.}(2021)\citenamefont {Ha},
  \citenamefont {Gupta}, \citenamefont {Rana},\ and\ \citenamefont
  {Venkatesh}}]{ha2021high}%
  \BibitemOpen
  \bibfield  {author} {\bibinfo {author} {\bibfnamefont {H.}~\bibnamefont
  {Ha}}, \bibinfo {author} {\bibfnamefont {S.}~\bibnamefont {Gupta}}, \bibinfo
  {author} {\bibfnamefont {S.}~\bibnamefont {Rana}},\ and\ \bibinfo {author}
  {\bibfnamefont {S.}~\bibnamefont {Venkatesh}},\ }\bibfield  {title} {\enquote
  {\bibinfo {title} {High dimensional level set estimation with bayesian neural
  network},}\ }in\ \href@noop {} {\emph {\bibinfo {booktitle} {Proceedings of
  the AAAI Conference on Artificial Intelligence}}},\ Vol.~\bibinfo {volume}
  {35}\ (\bibinfo {year} {2021})\ pp.\ \bibinfo {pages}
  {12095--12103}\BibitemShut {NoStop}%
\bibitem [{\citenamefont {Terayama}\ \emph {et~al.}(2019)\citenamefont
  {Terayama}, \citenamefont {Tamura}, \citenamefont {Nose}, \citenamefont
  {Hiramatsu}, \citenamefont {Hosono}, \citenamefont {Okuno},\ and\
  \citenamefont {Tsuda}}]{terayama2019efficient}%
  \BibitemOpen
  \bibfield  {author} {\bibinfo {author} {\bibfnamefont {K.}~\bibnamefont
  {Terayama}}, \bibinfo {author} {\bibfnamefont {R.}~\bibnamefont {Tamura}},
  \bibinfo {author} {\bibfnamefont {Y.}~\bibnamefont {Nose}}, \bibinfo {author}
  {\bibfnamefont {H.}~\bibnamefont {Hiramatsu}}, \bibinfo {author}
  {\bibfnamefont {H.}~\bibnamefont {Hosono}}, \bibinfo {author} {\bibfnamefont
  {Y.}~\bibnamefont {Okuno}},\ and\ \bibinfo {author} {\bibfnamefont
  {K.}~\bibnamefont {Tsuda}},\ }\bibfield  {title} {\enquote {\bibinfo {title}
  {Efficient construction method for phase diagrams using uncertainty
  sampling},}\ }\href@noop {} {\bibfield  {journal} {\bibinfo  {journal}
  {Physical Review Materials}\ }\textbf {\bibinfo {volume} {3}},\ \bibinfo
  {pages} {033802} (\bibinfo {year} {2019})}\BibitemShut {NoStop}%
\bibitem [{\citenamefont {Dai}\ and\ \citenamefont
  {Glotzer}(2020)}]{dai2020efficient}%
  \BibitemOpen
  \bibfield  {author} {\bibinfo {author} {\bibfnamefont {C.}~\bibnamefont
  {Dai}}\ and\ \bibinfo {author} {\bibfnamefont {S.~C.}\ \bibnamefont
  {Glotzer}},\ }\bibfield  {title} {\enquote {\bibinfo {title} {Efficient phase
  diagram sampling by active learning},}\ }\href@noop {} {\bibfield  {journal}
  {\bibinfo  {journal} {The Journal of Physical Chemistry B}\ }\textbf
  {\bibinfo {volume} {124}},\ \bibinfo {pages} {1275--1284} (\bibinfo {year}
  {2020})}\BibitemShut {NoStop}%
\bibitem [{\citenamefont {Fong}\ \emph {et~al.}(2021)\citenamefont {Fong},
  \citenamefont {Pellouchoud}, \citenamefont {Davidson}, \citenamefont
  {Walroth}, \citenamefont {Church}, \citenamefont {Tcareva}, \citenamefont
  {Wu}, \citenamefont {Peterson}, \citenamefont {Meredig},\ and\ \citenamefont
  {Tassone}}]{fong2021utilization}%
  \BibitemOpen
  \bibfield  {author} {\bibinfo {author} {\bibfnamefont {A.~Y.}\ \bibnamefont
  {Fong}}, \bibinfo {author} {\bibfnamefont {L.}~\bibnamefont {Pellouchoud}},
  \bibinfo {author} {\bibfnamefont {M.}~\bibnamefont {Davidson}}, \bibinfo
  {author} {\bibfnamefont {R.~C.}\ \bibnamefont {Walroth}}, \bibinfo {author}
  {\bibfnamefont {C.}~\bibnamefont {Church}}, \bibinfo {author} {\bibfnamefont
  {E.}~\bibnamefont {Tcareva}}, \bibinfo {author} {\bibfnamefont
  {L.}~\bibnamefont {Wu}}, \bibinfo {author} {\bibfnamefont {K.}~\bibnamefont
  {Peterson}}, \bibinfo {author} {\bibfnamefont {B.}~\bibnamefont {Meredig}},\
  and\ \bibinfo {author} {\bibfnamefont {C.~J.}\ \bibnamefont {Tassone}},\
  }\bibfield  {title} {\enquote {\bibinfo {title} {Utilization of machine
  learning to accelerate colloidal synthesis and discovery},}\ }\href@noop {}
  {\bibfield  {journal} {\bibinfo  {journal} {The Journal of Chemical Physics}\
  }\textbf {\bibinfo {volume} {154}},\ \bibinfo {pages} {224201} (\bibinfo
  {year} {2021})}\BibitemShut {NoStop}%
\bibitem [{\citenamefont {Feng}\ \emph {et~al.}(2020)\citenamefont {Feng},
  \citenamefont {Zelaya}, \citenamefont {Holm}, \citenamefont {Yang},\ and\
  \citenamefont {Cargnello}}]{feng2020investigation}%
  \BibitemOpen
  \bibfield  {author} {\bibinfo {author} {\bibfnamefont {E.~Y.}\ \bibnamefont
  {Feng}}, \bibinfo {author} {\bibfnamefont {R.}~\bibnamefont {Zelaya}},
  \bibinfo {author} {\bibfnamefont {A.}~\bibnamefont {Holm}}, \bibinfo {author}
  {\bibfnamefont {A.-C.}\ \bibnamefont {Yang}},\ and\ \bibinfo {author}
  {\bibfnamefont {M.}~\bibnamefont {Cargnello}},\ }\bibfield  {title} {\enquote
  {\bibinfo {title} {Investigation of the optical properties of uniform
  platinum, palladium, and nickel nanocrystals enables direct measurements of
  their concentrations in solution},}\ }\href@noop {} {\bibfield  {journal}
  {\bibinfo  {journal} {Colloids and Surfaces A: Physicochemical and
  Engineering Aspects}\ }\textbf {\bibinfo {volume} {601}},\ \bibinfo {pages}
  {125007} (\bibinfo {year} {2020})}\BibitemShut {NoStop}%
\bibitem [{\citenamefont {Rivnay}\ \emph {et~al.}(2016)\citenamefont {Rivnay},
  \citenamefont {Inal}, \citenamefont {Collins}, \citenamefont {Sessolo},
  \citenamefont {Stavrinidou}, \citenamefont {Strakosas}, \citenamefont
  {Tassone}, \citenamefont {Delongchamp},\ and\ \citenamefont
  {Malliaras}}]{rivnay2016structural}%
  \BibitemOpen
  \bibfield  {author} {\bibinfo {author} {\bibfnamefont {J.}~\bibnamefont
  {Rivnay}}, \bibinfo {author} {\bibfnamefont {S.}~\bibnamefont {Inal}},
  \bibinfo {author} {\bibfnamefont {B.~A.}\ \bibnamefont {Collins}}, \bibinfo
  {author} {\bibfnamefont {M.}~\bibnamefont {Sessolo}}, \bibinfo {author}
  {\bibfnamefont {E.}~\bibnamefont {Stavrinidou}}, \bibinfo {author}
  {\bibfnamefont {X.}~\bibnamefont {Strakosas}}, \bibinfo {author}
  {\bibfnamefont {C.}~\bibnamefont {Tassone}}, \bibinfo {author} {\bibfnamefont
  {D.~M.}\ \bibnamefont {Delongchamp}},\ and\ \bibinfo {author} {\bibfnamefont
  {G.~G.}\ \bibnamefont {Malliaras}},\ }\bibfield  {title} {\enquote {\bibinfo
  {title} {Structural control of mixed ionic and electronic transport in
  conducting polymers},}\ }\href@noop {} {\bibfield  {journal} {\bibinfo
  {journal} {Nature communications}\ }\textbf {\bibinfo {volume} {7}},\
  \bibinfo {pages} {11287} (\bibinfo {year} {2016})}\BibitemShut {NoStop}%
\bibitem [{\citenamefont {Prentiss}, \citenamefont {Wales},\ and\ \citenamefont
  {Wolynes}(2010)}]{prentiss2010energy}%
  \BibitemOpen
  \bibfield  {author} {\bibinfo {author} {\bibfnamefont {M.~C.}\ \bibnamefont
  {Prentiss}}, \bibinfo {author} {\bibfnamefont {D.~J.}\ \bibnamefont
  {Wales}},\ and\ \bibinfo {author} {\bibfnamefont {P.~G.}\ \bibnamefont
  {Wolynes}},\ }\bibfield  {title} {\enquote {\bibinfo {title} {The energy
  landscape, folding pathways and the kinetics of a knotted protein},}\
  }\href@noop {} {\bibfield  {journal} {\bibinfo  {journal} {PLoS computational
  biology}\ }\textbf {\bibinfo {volume} {6}},\ \bibinfo {pages} {e1000835}
  (\bibinfo {year} {2010})}\BibitemShut {NoStop}%
\bibitem [{\citenamefont {Singh}\ \emph {et~al.}(2018)\citenamefont {Singh},
  \citenamefont {Montoya}, \citenamefont {Rohr}, \citenamefont {Tsai},
  \citenamefont {Vojvodic},\ and\ \citenamefont
  {N{\o}rskov}}]{singh2018computational}%
  \BibitemOpen
  \bibfield  {author} {\bibinfo {author} {\bibfnamefont {A.~R.}\ \bibnamefont
  {Singh}}, \bibinfo {author} {\bibfnamefont {J.~H.}\ \bibnamefont {Montoya}},
  \bibinfo {author} {\bibfnamefont {B.~A.}\ \bibnamefont {Rohr}}, \bibinfo
  {author} {\bibfnamefont {C.}~\bibnamefont {Tsai}}, \bibinfo {author}
  {\bibfnamefont {A.}~\bibnamefont {Vojvodic}},\ and\ \bibinfo {author}
  {\bibfnamefont {J.~K.}\ \bibnamefont {N{\o}rskov}},\ }\bibfield  {title}
  {\enquote {\bibinfo {title} {Computational design of active site structures
  with improved transition-state scaling for ammonia synthesis},}\ }\href@noop
  {} {\bibfield  {journal} {\bibinfo  {journal} {ACS Catalysis}\ }\textbf
  {\bibinfo {volume} {8}},\ \bibinfo {pages} {4017--4024} (\bibinfo {year}
  {2018})}\BibitemShut {NoStop}%
\bibitem [{\citenamefont {Foloppe}\ \emph {et~al.}(2006)\citenamefont
  {Foloppe}, \citenamefont {Fisher}, \citenamefont {Howes}, \citenamefont
  {Potter}, \citenamefont {Robertson},\ and\ \citenamefont
  {Surgenor}}]{foloppe2006identification}%
  \BibitemOpen
  \bibfield  {author} {\bibinfo {author} {\bibfnamefont {N.}~\bibnamefont
  {Foloppe}}, \bibinfo {author} {\bibfnamefont {L.~M.}\ \bibnamefont {Fisher}},
  \bibinfo {author} {\bibfnamefont {R.}~\bibnamefont {Howes}}, \bibinfo
  {author} {\bibfnamefont {A.}~\bibnamefont {Potter}}, \bibinfo {author}
  {\bibfnamefont {A.~G.}\ \bibnamefont {Robertson}},\ and\ \bibinfo {author}
  {\bibfnamefont {A.~E.}\ \bibnamefont {Surgenor}},\ }\bibfield  {title}
  {\enquote {\bibinfo {title} {Identification of chemically diverse chk1
  inhibitors by receptor-based virtual screening},}\ }\href@noop {} {\bibfield
  {journal} {\bibinfo  {journal} {Bioorganic \& medicinal chemistry}\ }\textbf
  {\bibinfo {volume} {14}},\ \bibinfo {pages} {4792--4802} (\bibinfo {year}
  {2006})}\BibitemShut {NoStop}%
\bibitem [{\citenamefont {Palac{\'\i}n}\ and\ \citenamefont
  {de~Guibert}(2016)}]{palacin2016batteries}%
  \BibitemOpen
  \bibfield  {author} {\bibinfo {author} {\bibfnamefont {M.~R.}\ \bibnamefont
  {Palac{\'\i}n}}\ and\ \bibinfo {author} {\bibfnamefont {A.}~\bibnamefont
  {de~Guibert}},\ }\bibfield  {title} {\enquote {\bibinfo {title} {Why do
  batteries fail?}}\ }\href@noop {} {\bibfield  {journal} {\bibinfo  {journal}
  {Science}\ }\textbf {\bibinfo {volume} {351}},\ \bibinfo {pages} {1253292}
  (\bibinfo {year} {2016})}\BibitemShut {NoStop}%
\bibitem [{\citenamefont {Scott}(2018)}]{scott2018matter}%
  \BibitemOpen
  \bibfield  {author} {\bibinfo {author} {\bibfnamefont {S.~L.}\ \bibnamefont
  {Scott}},\ }\href@noop {} {\enquote {\bibinfo {title} {A matter of life
  (time) and death},}\ } (\bibinfo {year} {2018})\BibitemShut {NoStop}%
\bibitem [{\citenamefont {J{\o}rgensen}, \citenamefont {Norrman},\ and\
  \citenamefont {Krebs}(2008)}]{jorgensen2008stability}%
  \BibitemOpen
  \bibfield  {author} {\bibinfo {author} {\bibfnamefont {M.}~\bibnamefont
  {J{\o}rgensen}}, \bibinfo {author} {\bibfnamefont {K.}~\bibnamefont
  {Norrman}},\ and\ \bibinfo {author} {\bibfnamefont {F.~C.}\ \bibnamefont
  {Krebs}},\ }\bibfield  {title} {\enquote {\bibinfo {title}
  {Stability/degradation of polymer solar cells},}\ }\href@noop {} {\bibfield
  {journal} {\bibinfo  {journal} {Solar energy materials and solar cells}\
  }\textbf {\bibinfo {volume} {92}},\ \bibinfo {pages} {686--714} (\bibinfo
  {year} {2008})}\BibitemShut {NoStop}%
\bibitem [{\citenamefont {Di}\ and\ \citenamefont {Kerns}(2015)}]{di2015drug}%
  \BibitemOpen
  \bibfield  {author} {\bibinfo {author} {\bibfnamefont {L.}~\bibnamefont
  {Di}}\ and\ \bibinfo {author} {\bibfnamefont {E.~H.}\ \bibnamefont {Kerns}},\
  }\href@noop {} {\emph {\bibinfo {title} {Drug-like properties: concepts,
  structure design and methods from ADME to toxicity optimization}}}\ (\bibinfo
   {publisher} {Academic press},\ \bibinfo {year} {2015})\BibitemShut {NoStop}%
\bibitem [{\citenamefont {Neiswanger}, \citenamefont {Wang},\ and\
  \citenamefont {Ermon}(2021)}]{neiswanger2021bayesian}%
  \BibitemOpen
  \bibfield  {author} {\bibinfo {author} {\bibfnamefont {W.}~\bibnamefont
  {Neiswanger}}, \bibinfo {author} {\bibfnamefont {K.~A.}\ \bibnamefont
  {Wang}},\ and\ \bibinfo {author} {\bibfnamefont {S.}~\bibnamefont {Ermon}},\
  }\bibfield  {title} {\enquote {\bibinfo {title} {Bayesian algorithm
  execution: Estimating computable properties of black-box functions using
  mutual information},}\ }in\ \href@noop {} {\emph {\bibinfo {booktitle}
  {International Conference on Machine Learning}}}\ (\bibinfo {organization}
  {PMLR},\ \bibinfo {year} {2021})\ pp.\ \bibinfo {pages}
  {8005--8015}\BibitemShut {NoStop}%
\bibitem [{\citenamefont {Miskovich}\ \emph {et~al.}(2022)\citenamefont
  {Miskovich}, \citenamefont {Neiswanger}, \citenamefont {Colocho},
  \citenamefont {Emma}, \citenamefont {Garrahan}, \citenamefont {Maxwell},
  \citenamefont {Mayes}, \citenamefont {Ermon}, \citenamefont {Edelen},\ and\
  \citenamefont {Ratner}}]{miskovich2022bayesian}%
  \BibitemOpen
  \bibfield  {author} {\bibinfo {author} {\bibfnamefont {S.~A.}\ \bibnamefont
  {Miskovich}}, \bibinfo {author} {\bibfnamefont {W.}~\bibnamefont
  {Neiswanger}}, \bibinfo {author} {\bibfnamefont {W.}~\bibnamefont {Colocho}},
  \bibinfo {author} {\bibfnamefont {C.}~\bibnamefont {Emma}}, \bibinfo {author}
  {\bibfnamefont {J.}~\bibnamefont {Garrahan}}, \bibinfo {author}
  {\bibfnamefont {T.}~\bibnamefont {Maxwell}}, \bibinfo {author} {\bibfnamefont
  {C.}~\bibnamefont {Mayes}}, \bibinfo {author} {\bibfnamefont
  {S.}~\bibnamefont {Ermon}}, \bibinfo {author} {\bibfnamefont
  {A.}~\bibnamefont {Edelen}},\ and\ \bibinfo {author} {\bibfnamefont
  {D.}~\bibnamefont {Ratner}},\ }\bibfield  {title} {\enquote {\bibinfo {title}
  {Bayesian algorithm execution for tuning particle accelerator emittance with
  partial measurements},}\ }\href@noop {} {\bibfield  {journal} {\bibinfo
  {journal} {arXiv preprint arXiv:2209.04587}\ } (\bibinfo {year}
  {2022})}\BibitemShut {NoStop}%
\bibitem [{\citenamefont {Katsube}\ \emph {et~al.}(2020)\citenamefont
  {Katsube}, \citenamefont {Terayama}, \citenamefont {Tamura},\ and\
  \citenamefont {Nose}}]{katsube2020experimental}%
  \BibitemOpen
  \bibfield  {author} {\bibinfo {author} {\bibfnamefont {R.}~\bibnamefont
  {Katsube}}, \bibinfo {author} {\bibfnamefont {K.}~\bibnamefont {Terayama}},
  \bibinfo {author} {\bibfnamefont {R.}~\bibnamefont {Tamura}},\ and\ \bibinfo
  {author} {\bibfnamefont {Y.}~\bibnamefont {Nose}},\ }\bibfield  {title}
  {\enquote {\bibinfo {title} {Experimental establishment of phase diagrams
  guided by uncertainty sampling: an application to the deposition of zn--sn--p
  films by molecular beam epitaxy},}\ }\href@noop {} {\bibfield  {journal}
  {\bibinfo  {journal} {ACS Materials Letters}\ }\textbf {\bibinfo {volume}
  {2}},\ \bibinfo {pages} {571--575} (\bibinfo {year} {2020})}\BibitemShut
  {NoStop}%
\bibitem [{\citenamefont {Torres}\ \emph {et~al.}(2019)\citenamefont {Torres},
  \citenamefont {Jennings}, \citenamefont {Hansen}, \citenamefont {Boes},\ and\
  \citenamefont {Bligaard}}]{torres2019low}%
  \BibitemOpen
  \bibfield  {author} {\bibinfo {author} {\bibfnamefont {J.~A.~G.}\
  \bibnamefont {Torres}}, \bibinfo {author} {\bibfnamefont {P.~C.}\
  \bibnamefont {Jennings}}, \bibinfo {author} {\bibfnamefont {M.~H.}\
  \bibnamefont {Hansen}}, \bibinfo {author} {\bibfnamefont {J.~R.}\
  \bibnamefont {Boes}},\ and\ \bibinfo {author} {\bibfnamefont
  {T.}~\bibnamefont {Bligaard}},\ }\bibfield  {title} {\enquote {\bibinfo
  {title} {Low-scaling algorithm for nudged elastic band calculations using a
  surrogate machine learning model},}\ }\href@noop {} {\bibfield  {journal}
  {\bibinfo  {journal} {Physical review letters}\ }\textbf {\bibinfo {volume}
  {122}},\ \bibinfo {pages} {156001} (\bibinfo {year} {2019})}\BibitemShut
  {NoStop}%
\bibitem [{\citenamefont {Tian}\ \emph {et~al.}(2021)\citenamefont {Tian},
  \citenamefont {Yuan}, \citenamefont {Xue}, \citenamefont {Zhou},
  \citenamefont {Wang}, \citenamefont {Ding}, \citenamefont {Sun},\ and\
  \citenamefont {Lookman}}]{tian2021determining}%
  \BibitemOpen
  \bibfield  {author} {\bibinfo {author} {\bibfnamefont {Y.}~\bibnamefont
  {Tian}}, \bibinfo {author} {\bibfnamefont {R.}~\bibnamefont {Yuan}}, \bibinfo
  {author} {\bibfnamefont {D.}~\bibnamefont {Xue}}, \bibinfo {author}
  {\bibfnamefont {Y.}~\bibnamefont {Zhou}}, \bibinfo {author} {\bibfnamefont
  {Y.}~\bibnamefont {Wang}}, \bibinfo {author} {\bibfnamefont {X.}~\bibnamefont
  {Ding}}, \bibinfo {author} {\bibfnamefont {J.}~\bibnamefont {Sun}},\ and\
  \bibinfo {author} {\bibfnamefont {T.}~\bibnamefont {Lookman}},\ }\bibfield
  {title} {\enquote {\bibinfo {title} {Determining multi-component phase
  diagrams with desired characteristics using active learning},}\ }\href@noop
  {} {\bibfield  {journal} {\bibinfo  {journal} {Advanced Science}\ }\textbf
  {\bibinfo {volume} {8}},\ \bibinfo {pages} {2003165} (\bibinfo {year}
  {2021})}\BibitemShut {NoStop}%
\bibitem [{\citenamefont {Pellegrino}\ \emph {et~al.}(2020)\citenamefont
  {Pellegrino}, \citenamefont {Isopescu}, \citenamefont {Pelluti{\`e}},
  \citenamefont {Sordello}, \citenamefont {Rossi}, \citenamefont {Ortel},
  \citenamefont {Martra}, \citenamefont {Hodoroaba},\ and\ \citenamefont
  {Maurino}}]{pellegrino2020machine}%
  \BibitemOpen
  \bibfield  {author} {\bibinfo {author} {\bibfnamefont {F.}~\bibnamefont
  {Pellegrino}}, \bibinfo {author} {\bibfnamefont {R.}~\bibnamefont
  {Isopescu}}, \bibinfo {author} {\bibfnamefont {L.}~\bibnamefont
  {Pelluti{\`e}}}, \bibinfo {author} {\bibfnamefont {F.}~\bibnamefont
  {Sordello}}, \bibinfo {author} {\bibfnamefont {A.~M.}\ \bibnamefont {Rossi}},
  \bibinfo {author} {\bibfnamefont {E.}~\bibnamefont {Ortel}}, \bibinfo
  {author} {\bibfnamefont {G.}~\bibnamefont {Martra}}, \bibinfo {author}
  {\bibfnamefont {V.-D.}\ \bibnamefont {Hodoroaba}},\ and\ \bibinfo {author}
  {\bibfnamefont {V.}~\bibnamefont {Maurino}},\ }\bibfield  {title} {\enquote
  {\bibinfo {title} {Machine learning approach for elucidating and predicting
  the role of synthesis parameters on the shape and size of tio2
  nanoparticles},}\ }\href@noop {} {\bibfield  {journal} {\bibinfo  {journal}
  {Scientific Reports}\ }\textbf {\bibinfo {volume} {10}},\ \bibinfo {pages}
  {18910} (\bibinfo {year} {2020})}\BibitemShut {NoStop}%
\bibitem [{\citenamefont {Tassone}\ and\ \citenamefont
  {Mehta}(2019)}]{tassone2019aggregation}%
  \BibitemOpen
  \bibfield  {author} {\bibinfo {author} {\bibfnamefont {C.}~\bibnamefont
  {Tassone}}\ and\ \bibinfo {author} {\bibfnamefont {A.}~\bibnamefont
  {Mehta}},\ }\href@noop {} {\enquote {\bibinfo {title} {Aggregation and
  structuring of materials and chemicals data from diverse sources},}\
  }\bibinfo {type} {Tech. Rep.}\ (\bibinfo  {institution} {SLAC National
  Accelerator Lab., Menlo Park, CA (United States)},\ \bibinfo {year}
  {2019})\BibitemShut {NoStop}%
\bibitem [{\citenamefont {Yoo}\ \emph {et~al.}(2006)\citenamefont {Yoo},
  \citenamefont {Xue}, \citenamefont {Chu}, \citenamefont {Xu}, \citenamefont
  {Hangen}, \citenamefont {Lee}, \citenamefont {Stein},\ and\ \citenamefont
  {Xiang}}]{yoo2006identification}%
  \BibitemOpen
  \bibfield  {author} {\bibinfo {author} {\bibfnamefont {Y.~K.}\ \bibnamefont
  {Yoo}}, \bibinfo {author} {\bibfnamefont {Q.}~\bibnamefont {Xue}}, \bibinfo
  {author} {\bibfnamefont {Y.~S.}\ \bibnamefont {Chu}}, \bibinfo {author}
  {\bibfnamefont {S.}~\bibnamefont {Xu}}, \bibinfo {author} {\bibfnamefont
  {U.}~\bibnamefont {Hangen}}, \bibinfo {author} {\bibfnamefont {H.-C.}\
  \bibnamefont {Lee}}, \bibinfo {author} {\bibfnamefont {W.}~\bibnamefont
  {Stein}},\ and\ \bibinfo {author} {\bibfnamefont {X.-D.}\ \bibnamefont
  {Xiang}},\ }\bibfield  {title} {\enquote {\bibinfo {title} {Identification of
  amorphous phases in the fe--ni--co ternary alloy system using continuous
  phase diagram material chips},}\ }\href@noop {} {\bibfield  {journal}
  {\bibinfo  {journal} {Intermetallics}\ }\textbf {\bibinfo {volume} {14}},\
  \bibinfo {pages} {241--247} (\bibinfo {year} {2006})}\BibitemShut {NoStop}%
\bibitem [{\citenamefont {Antonov}\ \emph {et~al.}(1997)\citenamefont
  {Antonov}, \citenamefont {Oppeneer}, \citenamefont {Yaresko}, \citenamefont
  {Perlov},\ and\ \citenamefont {Kraft}}]{antonov1997computationally}%
  \BibitemOpen
  \bibfield  {author} {\bibinfo {author} {\bibfnamefont {V.}~\bibnamefont
  {Antonov}}, \bibinfo {author} {\bibfnamefont {P.}~\bibnamefont {Oppeneer}},
  \bibinfo {author} {\bibfnamefont {A.}~\bibnamefont {Yaresko}}, \bibinfo
  {author} {\bibfnamefont {A.~Y.}\ \bibnamefont {Perlov}},\ and\ \bibinfo
  {author} {\bibfnamefont {T.}~\bibnamefont {Kraft}},\ }\bibfield  {title}
  {\enquote {\bibinfo {title} {Computationally based explanation of the
  peculiar magneto-optical properties of ptmnsb and related ternary
  compounds},}\ }\href@noop {} {\bibfield  {journal} {\bibinfo  {journal}
  {Physical Review B}\ }\textbf {\bibinfo {volume} {56}},\ \bibinfo {pages}
  {13012} (\bibinfo {year} {1997})}\BibitemShut {NoStop}%
\bibitem [{\citenamefont {Abolhasani}\ and\ \citenamefont
  {Kumacheva}(2023)}]{abolhasani2023rise}%
  \BibitemOpen
  \bibfield  {author} {\bibinfo {author} {\bibfnamefont {M.}~\bibnamefont
  {Abolhasani}}\ and\ \bibinfo {author} {\bibfnamefont {E.}~\bibnamefont
  {Kumacheva}},\ }\bibfield  {title} {\enquote {\bibinfo {title} {The rise of
  self-driving labs in chemical and materials sciences},}\ }\href@noop {}
  {\bibfield  {journal} {\bibinfo  {journal} {Nature Synthesis}\ ,\ \bibinfo
  {pages} {1--10}} (\bibinfo {year} {2023})}\BibitemShut {NoStop}%
\bibitem [{\citenamefont {MacLeod}\ \emph {et~al.}(2020)\citenamefont
  {MacLeod}, \citenamefont {Parlane}, \citenamefont {Morrissey}, \citenamefont
  {H{\"a}se}, \citenamefont {Roch}, \citenamefont {Dettelbach}, \citenamefont
  {Moreira}, \citenamefont {Yunker}, \citenamefont {Rooney}, \citenamefont
  {Deeth} \emph {et~al.}}]{macleod2020self}%
  \BibitemOpen
  \bibfield  {author} {\bibinfo {author} {\bibfnamefont {B.~P.}\ \bibnamefont
  {MacLeod}}, \bibinfo {author} {\bibfnamefont {F.~G.}\ \bibnamefont
  {Parlane}}, \bibinfo {author} {\bibfnamefont {T.~D.}\ \bibnamefont
  {Morrissey}}, \bibinfo {author} {\bibfnamefont {F.}~\bibnamefont {H{\"a}se}},
  \bibinfo {author} {\bibfnamefont {L.~M.}\ \bibnamefont {Roch}}, \bibinfo
  {author} {\bibfnamefont {K.~E.}\ \bibnamefont {Dettelbach}}, \bibinfo
  {author} {\bibfnamefont {R.}~\bibnamefont {Moreira}}, \bibinfo {author}
  {\bibfnamefont {L.~P.}\ \bibnamefont {Yunker}}, \bibinfo {author}
  {\bibfnamefont {M.~B.}\ \bibnamefont {Rooney}}, \bibinfo {author}
  {\bibfnamefont {J.~R.}\ \bibnamefont {Deeth}}, \emph {et~al.},\ }\bibfield
  {title} {\enquote {\bibinfo {title} {Self-driving laboratory for accelerated
  discovery of thin-film materials},}\ }\href@noop {} {\bibfield  {journal}
  {\bibinfo  {journal} {Science Advances}\ }\textbf {\bibinfo {volume} {6}},\
  \bibinfo {pages} {eaaz8867} (\bibinfo {year} {2020})}\BibitemShut {NoStop}%
\bibitem [{\citenamefont {MacLeod}\ \emph {et~al.}(2022)\citenamefont
  {MacLeod}, \citenamefont {Parlane}, \citenamefont {Rupnow}, \citenamefont
  {Dettelbach}, \citenamefont {Elliott}, \citenamefont {Morrissey},
  \citenamefont {Haley}, \citenamefont {Proskurin}, \citenamefont {Rooney},
  \citenamefont {Taherimakhsousi} \emph {et~al.}}]{macleod2022self}%
  \BibitemOpen
  \bibfield  {author} {\bibinfo {author} {\bibfnamefont {B.~P.}\ \bibnamefont
  {MacLeod}}, \bibinfo {author} {\bibfnamefont {F.~G.}\ \bibnamefont
  {Parlane}}, \bibinfo {author} {\bibfnamefont {C.~C.}\ \bibnamefont {Rupnow}},
  \bibinfo {author} {\bibfnamefont {K.~E.}\ \bibnamefont {Dettelbach}},
  \bibinfo {author} {\bibfnamefont {M.~S.}\ \bibnamefont {Elliott}}, \bibinfo
  {author} {\bibfnamefont {T.~D.}\ \bibnamefont {Morrissey}}, \bibinfo {author}
  {\bibfnamefont {T.~H.}\ \bibnamefont {Haley}}, \bibinfo {author}
  {\bibfnamefont {O.}~\bibnamefont {Proskurin}}, \bibinfo {author}
  {\bibfnamefont {M.~B.}\ \bibnamefont {Rooney}}, \bibinfo {author}
  {\bibfnamefont {N.}~\bibnamefont {Taherimakhsousi}}, \emph {et~al.},\
  }\bibfield  {title} {\enquote {\bibinfo {title} {A self-driving laboratory
  advances the pareto front for material properties},}\ }\href@noop {}
  {\bibfield  {journal} {\bibinfo  {journal} {Nature communications}\ }\textbf
  {\bibinfo {volume} {13}},\ \bibinfo {pages} {995} (\bibinfo {year}
  {2022})}\BibitemShut {NoStop}%
\bibitem [{\citenamefont {Szymanski}\ \emph
  {et~al.}(2023{\natexlab{b}})\citenamefont {Szymanski}, \citenamefont {Rendy},
  \citenamefont {Fei}, \citenamefont {Kumar}, \citenamefont {He}, \citenamefont
  {Milsted}, \citenamefont {McDermott}, \citenamefont {Gallant}, \citenamefont
  {Cubuk}, \citenamefont {Merchant}, \citenamefont {Kim}, \citenamefont {Jain},
  \citenamefont {Bartel}, \citenamefont {Persson}, \citenamefont {Zeng},\ and\
  \citenamefont {Ceder}}]{Szymanski2023}%
  \BibitemOpen
  \bibfield  {author} {\bibinfo {author} {\bibfnamefont {N.~J.}\ \bibnamefont
  {Szymanski}}, \bibinfo {author} {\bibfnamefont {B.}~\bibnamefont {Rendy}},
  \bibinfo {author} {\bibfnamefont {Y.}~\bibnamefont {Fei}}, \bibinfo {author}
  {\bibfnamefont {R.~E.}\ \bibnamefont {Kumar}}, \bibinfo {author}
  {\bibfnamefont {T.}~\bibnamefont {He}}, \bibinfo {author} {\bibfnamefont
  {D.}~\bibnamefont {Milsted}}, \bibinfo {author} {\bibfnamefont {M.~J.}\
  \bibnamefont {McDermott}}, \bibinfo {author} {\bibfnamefont {M.}~\bibnamefont
  {Gallant}}, \bibinfo {author} {\bibfnamefont {E.~D.}\ \bibnamefont {Cubuk}},
  \bibinfo {author} {\bibfnamefont {A.}~\bibnamefont {Merchant}}, \bibinfo
  {author} {\bibfnamefont {H.}~\bibnamefont {Kim}}, \bibinfo {author}
  {\bibfnamefont {A.}~\bibnamefont {Jain}}, \bibinfo {author} {\bibfnamefont
  {C.~J.}\ \bibnamefont {Bartel}}, \bibinfo {author} {\bibfnamefont
  {K.}~\bibnamefont {Persson}}, \bibinfo {author} {\bibfnamefont
  {Y.}~\bibnamefont {Zeng}},\ and\ \bibinfo {author} {\bibfnamefont
  {G.}~\bibnamefont {Ceder}},\ }\bibfield  {title} {\enquote {\bibinfo {title}
  {An autonomous laboratory for the accelerated synthesis of novel
  materials},}\ }\href {https://doi.org/10.1038/s41586-023-06734-w} {\bibfield
  {journal} {\bibinfo  {journal} {Nature}\ } (\bibinfo {year}
  {2023}{\natexlab{b}}),\ 10.1038/s41586-023-06734-w}\BibitemShut {NoStop}%
\bibitem [{\citenamefont {Chitturi}, \citenamefont {Ramdas},\ and\
  \citenamefont {Neiswanger}(2023{\natexlab{a}})}]{chitturi2023}%
  \BibitemOpen
  \bibfield  {author} {\bibinfo {author} {\bibfnamefont {S.}~\bibnamefont
  {Chitturi}}, \bibinfo {author} {\bibfnamefont {A.}~\bibnamefont {Ramdas}},\
  and\ \bibinfo {author} {\bibfnamefont {W.}~\bibnamefont {Neiswanger}},\
  }\href {https://doi.org/10.5281/zenodo.10246330} {\enquote {\bibinfo {title}
  {src47/multibax-sklearn},}\ } (\bibinfo {year}
  {2023}{\natexlab{a}})\BibitemShut {NoStop}%
\bibitem [{\citenamefont {Chitturi}, \citenamefont {Ramdas},\ and\
  \citenamefont {Neiswanger}(2023{\natexlab{b}})}]{chitturi2023gpflow}%
  \BibitemOpen
  \bibfield  {author} {\bibinfo {author} {\bibfnamefont {S.}~\bibnamefont
  {Chitturi}}, \bibinfo {author} {\bibfnamefont {A.}~\bibnamefont {Ramdas}},\
  and\ \bibinfo {author} {\bibfnamefont {W.}~\bibnamefont {Neiswanger}},\
  }\href {https://doi.org/10.5281/zenodo.10222982} {\enquote {\bibinfo {title}
  {src47/materials-bax-gpflow: Paper submission},}\ } (\bibinfo {year}
  {2023}{\natexlab{b}})\BibitemShut {NoStop}%
\bibitem [{\citenamefont {Matthews}\ \emph {et~al.}(2017)\citenamefont
  {Matthews}, \citenamefont {{van der Wilk}}, \citenamefont {Nickson},
  \citenamefont {Fujii}, \citenamefont {{Boukouvalas}}, \citenamefont
  {{Le{\'o}n-Villagr{\'a}}}, \citenamefont {Ghahramani},\ and\ \citenamefont
  {Hensman}}]{GPflow2017}%
  \BibitemOpen
  \bibfield  {author} {\bibinfo {author} {\bibfnamefont {A.~G. d.~G.}\
  \bibnamefont {Matthews}}, \bibinfo {author} {\bibfnamefont {M.}~\bibnamefont
  {{van der Wilk}}}, \bibinfo {author} {\bibfnamefont {T.}~\bibnamefont
  {Nickson}}, \bibinfo {author} {\bibfnamefont {K.}~\bibnamefont {Fujii}},
  \bibinfo {author} {\bibfnamefont {A.}~\bibnamefont {{Boukouvalas}}}, \bibinfo
  {author} {\bibfnamefont {P.}~\bibnamefont {{Le{\'o}n-Villagr{\'a}}}},
  \bibinfo {author} {\bibfnamefont {Z.}~\bibnamefont {Ghahramani}},\ and\
  \bibinfo {author} {\bibfnamefont {J.}~\bibnamefont {Hensman}},\ }\bibfield
  {title} {\enquote {\bibinfo {title} {{{GP}flow: A {G}aussian process library
  using {T}ensor{F}low}},}\ }\href {http://jmlr.org/papers/v18/16-537.html}
  {\bibfield  {journal} {\bibinfo  {journal} {Journal of Machine Learning
  Research}\ }\textbf {\bibinfo {volume} {18}},\ \bibinfo {pages} {1--6}
  (\bibinfo {year} {2017})}\BibitemShut {NoStop}%
\bibitem [{\citenamefont {Picheny}\ \emph {et~al.}(2023)\citenamefont
  {Picheny}, \citenamefont {Berkeley}, \citenamefont {Moss}, \citenamefont
  {Stojic}, \citenamefont {Granta}, \citenamefont {Ober}, \citenamefont
  {Artemev}, \citenamefont {Ghani}, \citenamefont {Goodall}, \citenamefont
  {Paleyes}, \citenamefont {Vakili}, \citenamefont {Pascual-Diaz},
  \citenamefont {Markou}, \citenamefont {Qing}, \citenamefont {Loka},\ and\
  \citenamefont {Couckuyt}}]{trieste2023}%
  \BibitemOpen
  \bibfield  {author} {\bibinfo {author} {\bibfnamefont {V.}~\bibnamefont
  {Picheny}}, \bibinfo {author} {\bibfnamefont {J.}~\bibnamefont {Berkeley}},
  \bibinfo {author} {\bibfnamefont {H.~B.}\ \bibnamefont {Moss}}, \bibinfo
  {author} {\bibfnamefont {H.}~\bibnamefont {Stojic}}, \bibinfo {author}
  {\bibfnamefont {U.}~\bibnamefont {Granta}}, \bibinfo {author} {\bibfnamefont
  {S.~W.}\ \bibnamefont {Ober}}, \bibinfo {author} {\bibfnamefont
  {A.}~\bibnamefont {Artemev}}, \bibinfo {author} {\bibfnamefont
  {K.}~\bibnamefont {Ghani}}, \bibinfo {author} {\bibfnamefont
  {A.}~\bibnamefont {Goodall}}, \bibinfo {author} {\bibfnamefont
  {A.}~\bibnamefont {Paleyes}}, \bibinfo {author} {\bibfnamefont
  {S.}~\bibnamefont {Vakili}}, \bibinfo {author} {\bibfnamefont
  {S.}~\bibnamefont {Pascual-Diaz}}, \bibinfo {author} {\bibfnamefont
  {S.}~\bibnamefont {Markou}}, \bibinfo {author} {\bibfnamefont
  {J.}~\bibnamefont {Qing}}, \bibinfo {author} {\bibfnamefont {N.~R. B.~S.}\
  \bibnamefont {Loka}},\ and\ \bibinfo {author} {\bibfnamefont
  {I.}~\bibnamefont {Couckuyt}},\ }\href
  {https://doi.org/10.48550/ARXIV.2302.08436} {\enquote {\bibinfo {title}
  {Trieste: Efficiently exploring the depths of black-box functions with
  tensorflow},}\ } (\bibinfo {year} {2023})\BibitemShut {NoStop}%
\bibitem [{\citenamefont {et~al}()}]{pythonternary}%
  \BibitemOpen
  \bibfield  {author} {\bibinfo {author} {\bibfnamefont {M.~H.}\ \bibnamefont
  {et~al}},\ }\bibfield  {title} {\enquote {\bibinfo {title} {python-ternary:
  Ternary plots in python},}\ }\href {https://doi.org/10.5281/zenodo.594435}
  {\bibfield  {journal} {\bibinfo  {journal} {Zenodo 10.5281/zenodo.594435}\
  }10.5281/zenodo.594435}\BibitemShut {NoStop}%
\end{thebibliography}%

\clearpage
\section*{Supplementary Text}
\setcounter{figure}{0}
\renewcommand{\figurename}{FIG.}
\renewcommand\thesubsection{\Alph{subsection}}
\renewcommand\thesubsubsection{\alph{subsubsection}}
\renewcommand{\thefigure}{S\arabic{figure}}

\subsection{Additional Algorithms for Materials Subset Estimation}
\label{sec:additional_algorithms}

In this section, we briefly describe two other useful flavors of algorithms developed here for searching materials spaces that are enabled by the BAX framework: percentile and conditional algorithms. 

\subsubsection{Percentile Algorithms}

\begin{figure}[p]
\centering
\includegraphics[width=\textwidth]{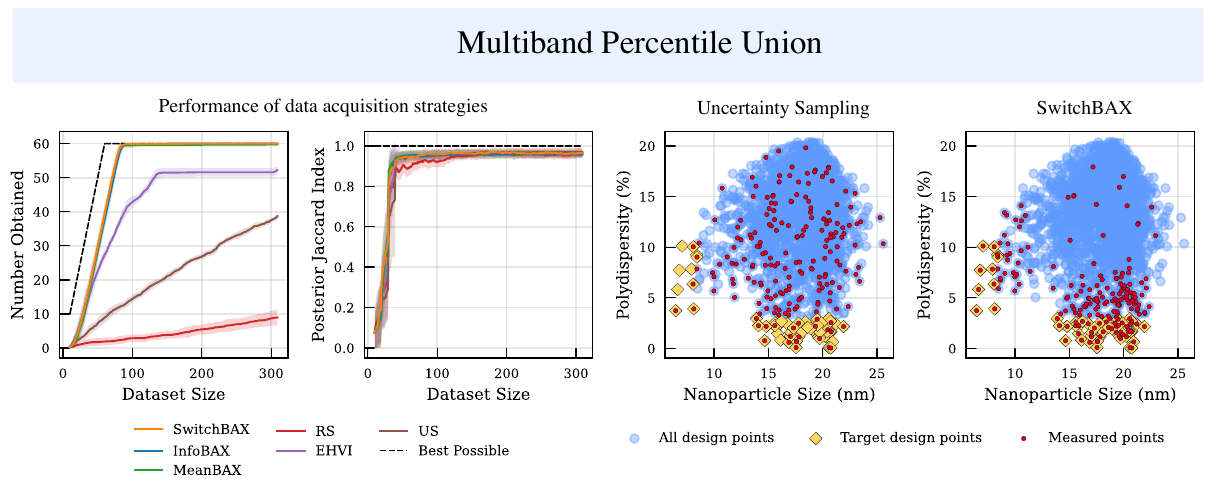}
\caption{Designing an acquisition strategy for the specific goal of finding nanoparticles with the lowest $0.5 \%$ radii values \textbf{or} the lowest $2.5 \%$ polydispersity percentage; percentage thresholds were chosen arbitrarily. Acquisition using this type of algorithm bypasses needing to specify explicit property thresholds. Note, the union operation (i.e. the `or' logic) was needed in this example as the intersection of individual percentile sets is empty.}
\label{fig:percentile_union}
\end{figure}

Percentile algorithms are distinct from algorithms which use pre-defined thresholds or bounds for measured properties. In this setting, the thresholds are not fixed (by value) explicitly.  Such algorithms are useful when it is unknown what specific thresholds should be chosen for a design space. For example, instead of stating the goal: ``Find all materials in the design space with measured property value greater than 5.0'', it is possible to ask ``Find the subset of the design space corresponding to the top 10$\%$ of measured property values''. For a design space with 100 material candidates, for case 1, it is possible that the target subset is empty. However, under case 2, the size of the target subset is exactly equal to 10 (for a single-property case). As an example, a possible percentile algorithm is to return the subset of the design space corresponding to the top $t_1\%$ of measured property 1 and/or the top $t_2\%$ of measured property 2 (Algorithms \ref{alg:PercentileIntersection}-\ref{alg:PercentileUnion}). An example of data acquisition targeting a percentile union is shown in Figure \ref{fig:percentile_union}.

\begin{algorithm}[H]
\caption{\textsc{MultiPercentileIntersection}: Determine input regions corresponding to the intersection of two percentile bands}
\begin{algorithmic}
\Function{MultiPercentileIntersection}{$f$, $X$} 
    \State $\mathcal{T}^x_1 \gets \text{argwhere}(f(X)_1 \geq \text{percentile}(f(X)_1, t_1))$ \algorithmiccomment{\textit{Property 1 percentile threshold}}
    \State $\mathcal{T}^x_2 \gets \text{argwhere}(f(X)_2 \geq \text{percentile}(f(X)_2, t_2))$ \algorithmiccomment{\textit{Property 2 percentile threshold}}
    \State $\mathcal{T} \gets \{ \mathcal{T}^x_1 \cap \mathcal{T}^x_2, f(\mathcal{T}^x_1 \cap \mathcal{T}^x_2) \}$ \algorithmiccomment{\textit{Intersection of percentile sets}}
    \State \Return $\mathcal{T}$
\EndFunction
\end{algorithmic}
\label{alg:PercentileIntersection}
\end{algorithm}

\begin{algorithm}[H]
\caption{\textsc{MultiPercentileUnion}: Union of two percentile bands}
\begin{algorithmic}
\Function{MultiPercentileUnion}{$f$, $X$} 
    \State $\mathcal{T}^x_1 \gets \text{argwhere}(f(X)_1 \geq \text{percentile}(f(X)_1, t_1))$ \algorithmiccomment{\textit{Property 1 percentile threshold}}
    \State $\mathcal{T}^x_2 \gets \text{argwhere}(f(X)_2 \geq \text{percentile}(f(X)_2, t_2))$ \algorithmiccomment{\textit{Property 2 percentile threshold}}
    \State $\mathcal{T} = \{ \mathcal{T}^x_1 \cup \mathcal{T}^x_2, f(\mathcal{T}^x_1 \cup \mathcal{T}^x_2) \}$ \algorithmiccomment{\textit{Union of percentile sets}}
    \State \Return $\mathcal{T}$
\EndFunction
\end{algorithmic}
\label{alg:PercentileUnion}
\end{algorithm}

Such a formalism is easily extended to a larger number of measured properties. However, one subtlety is that when there are more than two measured properties, there are multiple ways to combine the intersection and union operations. The BAX framework is agnostic to these user-choices. As long as subset filtering logic can be written, the corresponding request can be automatically converted into an acquisition function. 

\subsubsection{Conditional Algorithms}

One specific limitation of intersection algorithms (both the Multiband and Percentile algorithms) is that there may not be any points in the design space which jointly satisfy \textbf{both} band or percentile thresholds. For example, this could happen if the set of points corresponding to property 1 and 2 were disjoint. Prior to experimentation, it is typically unknown whether this set will be empty or not. We introduce a strategy based on conditional logic for increased robustness of data acquisition. Below, we describe a conditional algorithm which returns a multiband (intersection of two level bands) if it exists and otherwise returns target points corresponding to a level band in only one of the properties.

\begin{algorithm}[H]
\caption{\textsc{ConditionalMultiband}: Find level band if multiband is not achievable}
\begin{algorithmic}
\Function{ConditionalMultiband}{$f$, $X$} 
    \State $\mathcal{T}^x_1 \gets \text{argwhere}(a \leq f(X)_1 \leq b)$
    \State $\mathcal{T}^x_2 \gets \text{argwhere}(c \leq f(X)_2 \leq d)$ 
    \If {$\mathcal{T}^x_1 \cap \mathcal{T}^x_2 \neq \emptyset$} \algorithmiccomment{\textit{Check if multiband set is non-empty}}
        \State $\mathcal{T} = \{ \mathcal{T}^x_1 \cap \mathcal{T}^x_2, f(\mathcal{T}^x_1 \cap \mathcal{T}^x_2) \}$ \algorithmiccomment{\textit{Multiband in Property 1 and 2}}
        % \State \Return $\mathcal{T}$
    \Else
        \State $\mathcal{T} = \{\mathcal{T}^x_2, f(\mathcal{T}^x_2) \}$ \algorithmiccomment{\textit{Level band in Property 2}}
    \EndIf
    \State \Return $\mathcal{T}$
\EndFunction
\end{algorithmic}
\label{algo:ConditionalMultiband}
\end{algorithm}

\begin{figure}[p]
\centering
\includegraphics[width=0.9\textwidth]{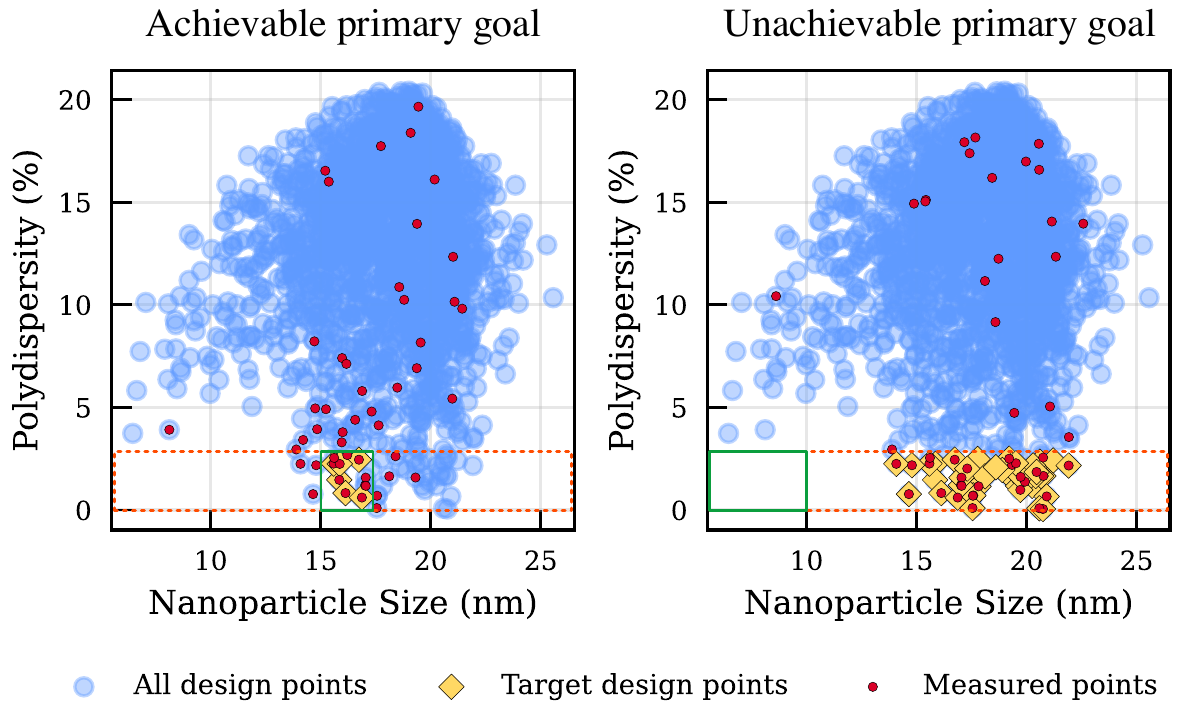}
\caption{Visualization of a Conditional Multiband Algorithm in (\textbf{A}) a case where the primary objective (green multiband) is achievable and (\textbf{B}) where the primary objective is unachievable and the algorithm output corresponds to the secondary objective (orange level band). Note, an objective being achievable means that there exists a least one point (shown in blue) in measured property space that falls within the green box. Sampling results are shown after 50 iterations. It is apparent that BAX strategies are able to correctly sample for goals based on conditional logic.}
\label{fig:conditional}
\end{figure}

Note, an algorithm executing on the ground-truth function will always yield the same output (i.e. the intersection either exists or it doesn't). However, when executing and algorithm on the posterior mean or draws from the trained \GP{} model, either condition may be achieved. Data acquisition is then suggested in a manner which aids in the determination of which condition actually holds. As a concrete example: one could specify the strategy of finding monodisperse 4nm particles if they are achievable given ranges on synthesis parameters and otherwise simply isolating conditions corresponding to monodisperse particles of any size. An example of a Conditional Multiband algorithm where the primary goal is achievable and unachievable is shown in Figure \ref{fig:conditional}. Crafting an algorithm in this way allows a user to bake in a fail-safe in situations where the experimental goal may be unachievable. This conditional logic can be extended to a hierarchy of conditions and is the subject of future work. 

\subsection{Comparison Against an Acquisition Function Designed for a Specific Task}
\label{sec:EHVI_comparison}

\begin{figure}[p]
\centering
\includegraphics[width=\textwidth]{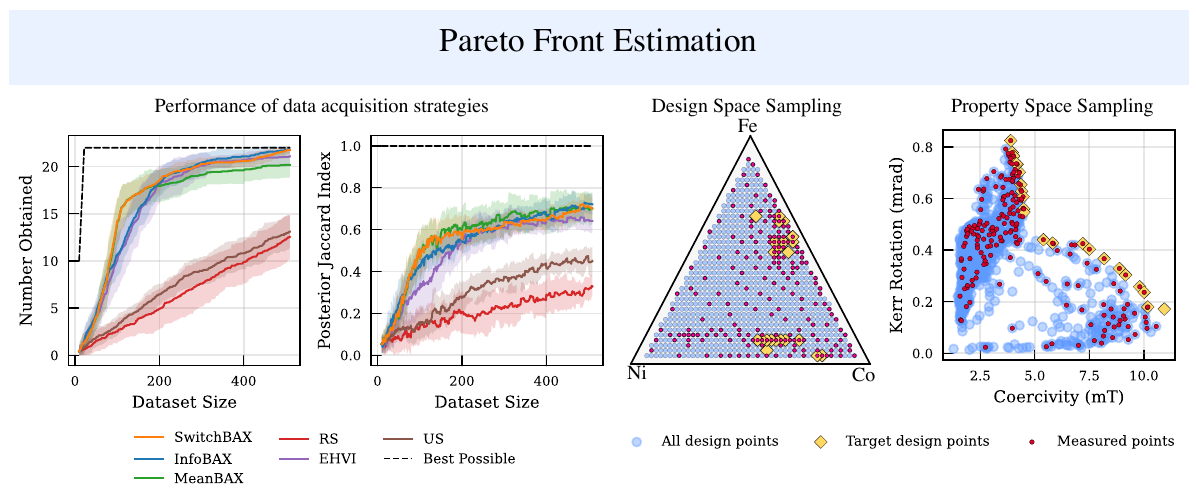}
\caption{Comparison between BAX strategies and an acquisition function (EHVI) designed for the specific goal of Pareto front optimization. BAX strategies compete favorably with EHVI for both the \nobtained{} and \Jposterior{} metrics, indicating that the algorithm-execution approach is not necessarily inferior to a detailed and complex custom acquisition function.}
\label{fig:paretofront}
\end{figure}

In general, the task of acquisition function development is challenging and quite technical. The advantage of the BAX approach is that a significant portion of this development is abstracted from the user. However, in this section, we sought to compare the performance of the BAX strategies on a multi-property task for which optimized acquisition functions already exist. We chose the task of Pareto front estimation (i.e. find the ground-truth target subset which corresponds to optimal trade offs in the measured properties). For this task, EHVI is one of several state-of-the-art algorithms. In Figure \ref{fig:paretofront}, we compare EHVI against MeanBAX, InfoBAX, and SwitchBAX and find that all approaches give similar long-term performance; interestingly, MeanBAX, and SwitchBAX actually outperform EHVI at low dataset sizes for this specific dataset. This important result shows that there is not necessarily a disadvantage in using a BAX strategy (in terms of the quality of collected samples), relative to a custom acquisition function. However, it is important to note that EHVI is a well optimized and differentiable acquisition function which is likely faster and more computationally efficient for large datasets or high-dimensional optimization. 

\subsection{Data Acquisition Using a Gaussian Process model with fixed hyperparameters}
\label{sec:fixed_hypers}

This section considers the algorithms presented in the main text (Library, Multiband, and Wishlist) under conditions of a \GP{} model with pre-fixed hyperparameters. Note, in the main text, results were presented using hyperparameters which were fit adaptively. While the adaptive hyperfitting scheme represents a more practical data collection scenario, it somewhat confounds the direct comparison between acquisition functions. Under an adaptive scheme, the model hyperparameters at any given iteration depends on the specific set of data collected. This means that the difference in performance between different sampling strategies could be ascribable to the combination of the acquisition function and the model, rather than just the acquisition function. 

\begin{figure}[p]
\centering
\includegraphics[width=\textwidth]{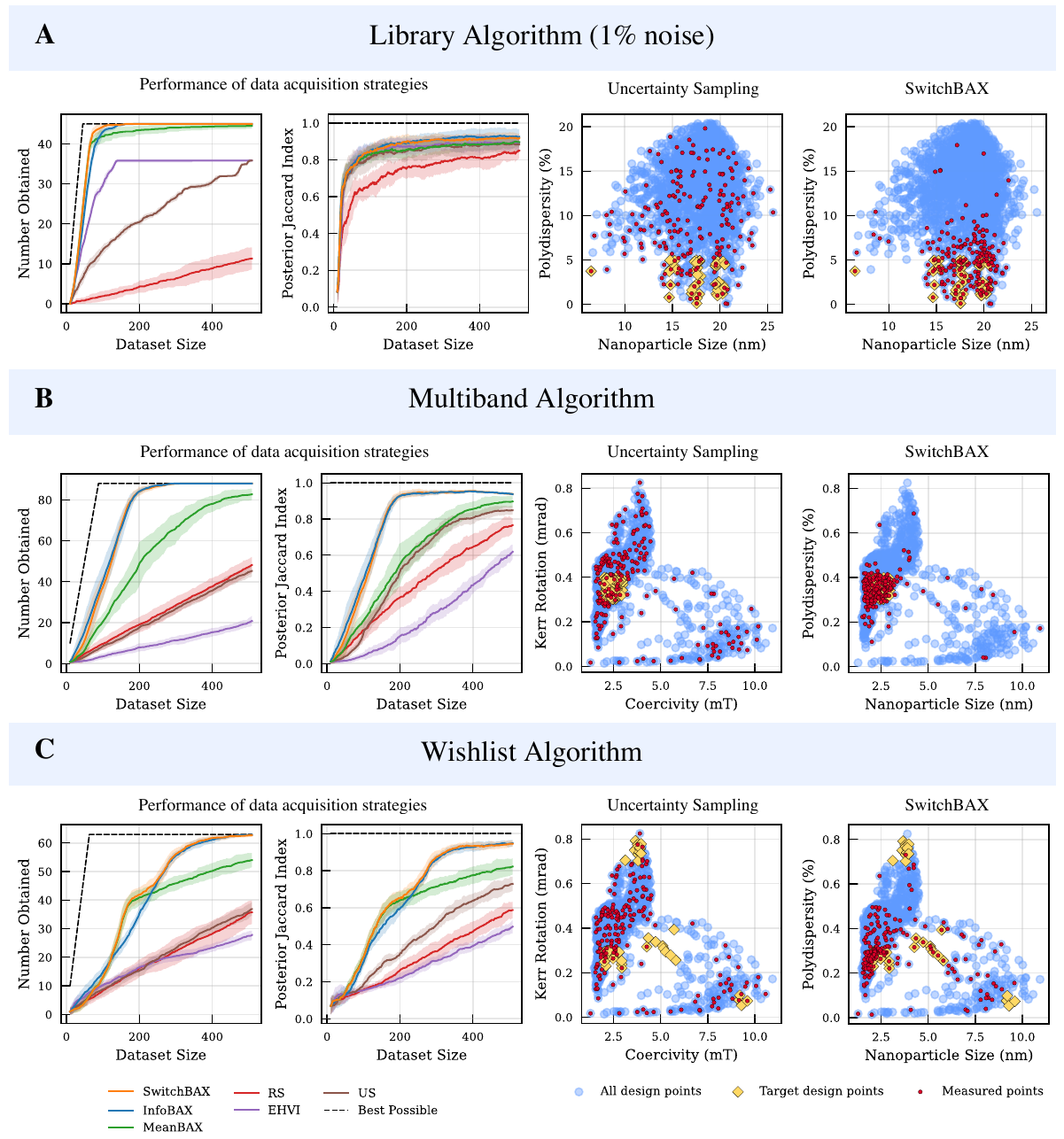}
\caption{Data acquisition corresponding to the (\textbf{A}) library example from
Figure 4 and (\textbf{B,C}) the Multiband and Wishlist algorithms in Figure 5 of the main text under the case where \GP{} hyperparameters are fixed throughout data acquisition. BAX strategies again prove to be the most efficient, with SwitchBAX maintaining good performance across all data regimes.}
\label{fig:fixedhypers}
\end{figure}

Therefore, we consider an idealized scenario in which we instead fit the \GP{} hyperparameters using 5 fold cross validation on the entire dataset and pick the hyperparameters corresponding to the largest log likelihood. These hyperparameters are fixed throughout data collection, allowing for a direct comparison between acquisition functions. The results for the three subset algorithms are shown in Figures \ref{fig:fixedhypers}A-C. We observe similar behavior to the adaptive hyperfitting case, with the goal-aware BAX strategies outperforming RS, US and EHVI. 

\clearpage
\section*{Supplementary Figures}

\begin{figure}[H]
\centering
\includegraphics[width=\textwidth]{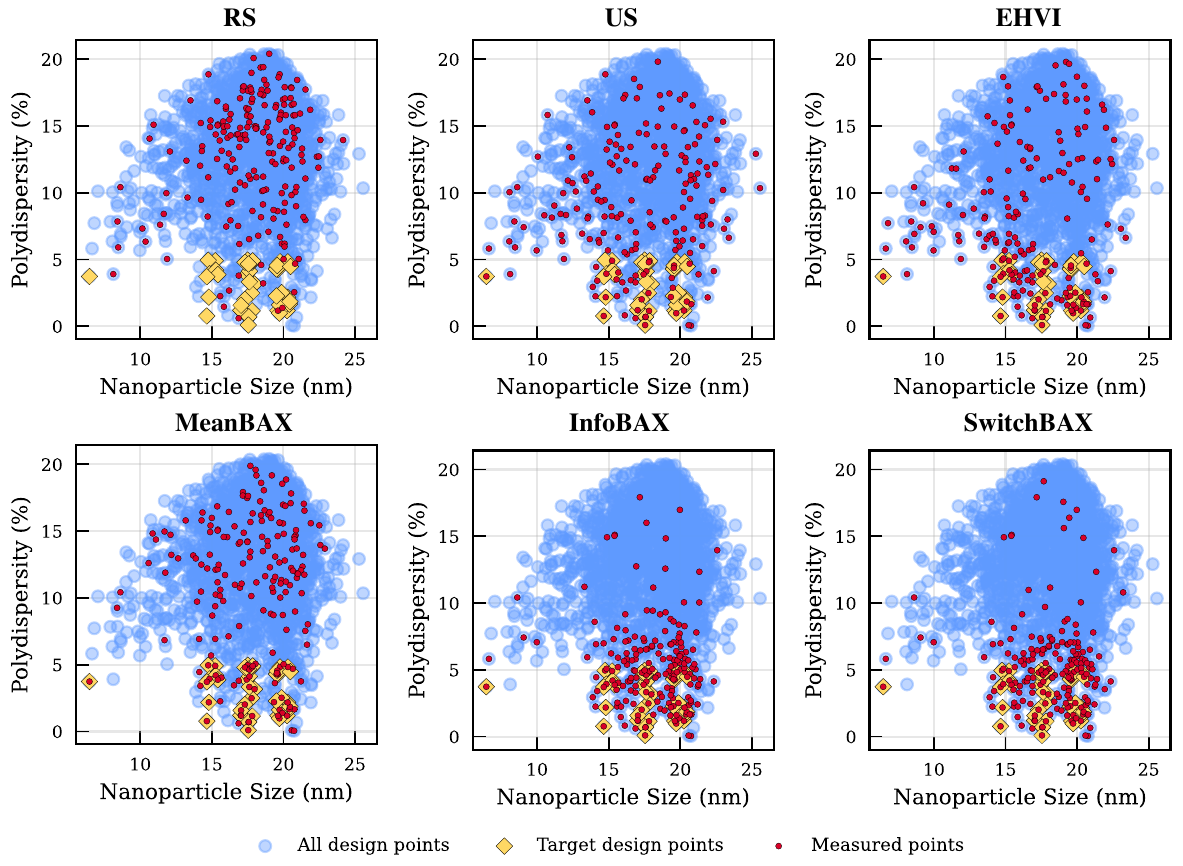}
\caption{Sampling results in measured property space after 200 iterations for the experimental goal of finding a library of monodisperse nanoparticles corresponding to Figure 4 in the main text. RS and US are relatively sample inefficient relative to the BAX strategies. EHVI seeks to find a Pareto front of minimal radius and polydispersity, which only has partial intersection with the ground truth target subset.}
\label{fig:np_y_sampling}
\end{figure}

\clearpage

\begin{figure}[p]
\centering
\includegraphics[width=0.9\textwidth]{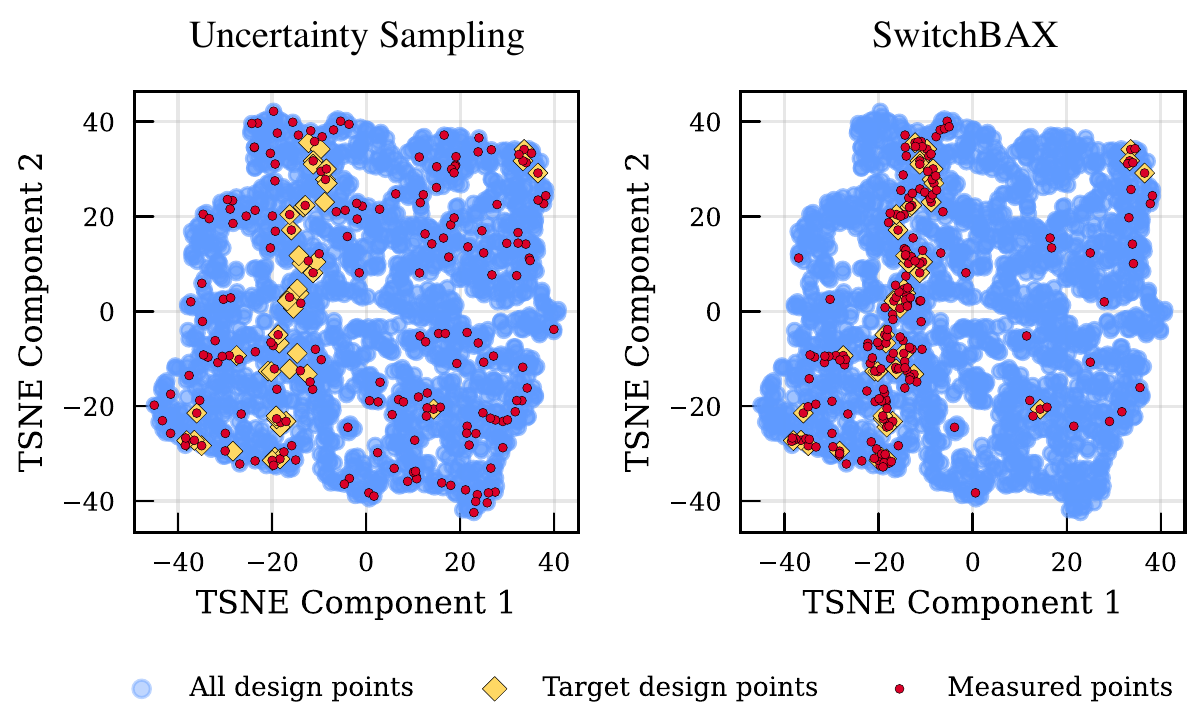}
\caption{Visualization of design space sampling for the experimental goal of finding a library of monodisperse nanoparticles (US and SwitchBAX). The four dimensional design space is represented here using t-distributed stochastic neighbor embedding (TSNE).}
\label{fig:np_x_sampling}
\end{figure}

\clearpage

\begin{figure}[p]
\centering
\includegraphics[width=\textwidth]{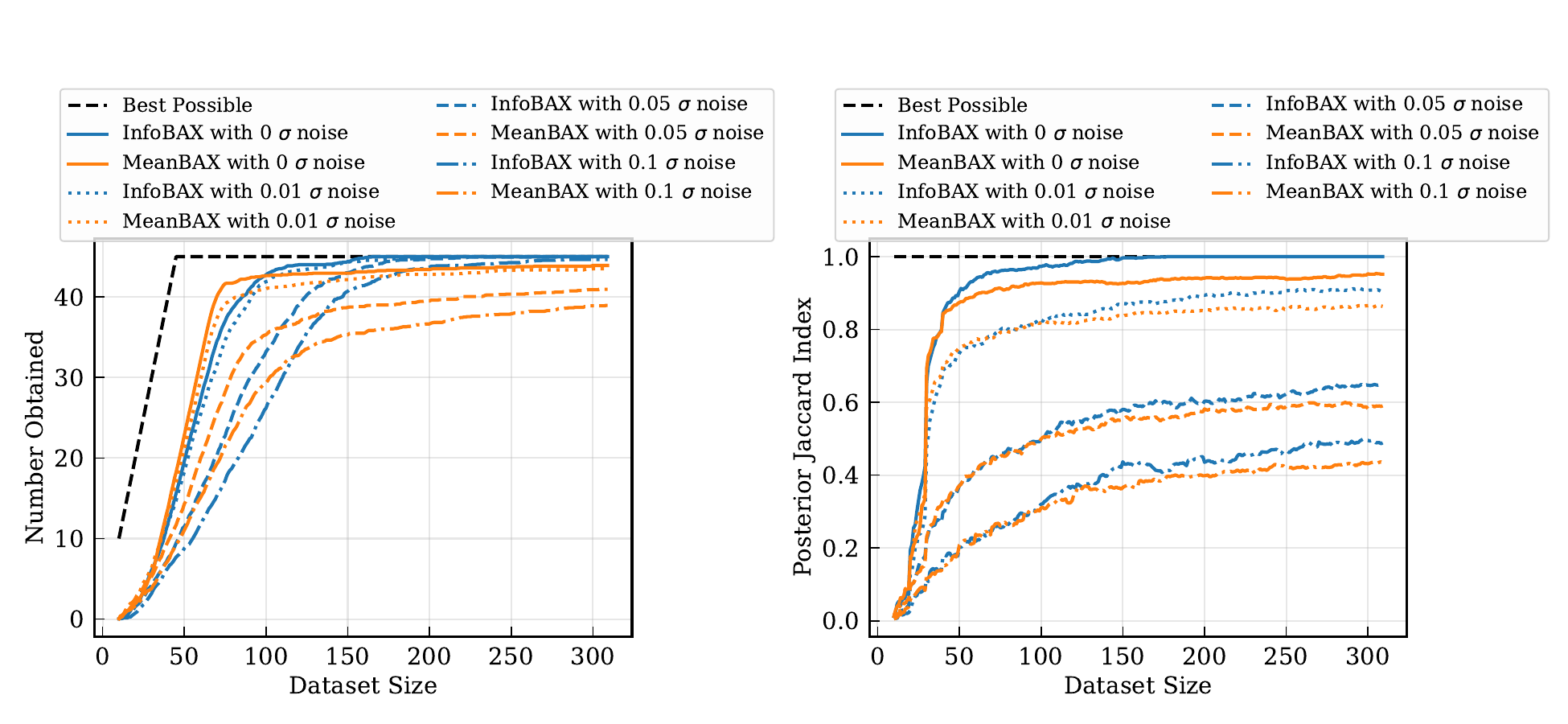}
\caption{Influence of noisy measurements ($\sigma$ = [0.0, 0.01, 0.05, 0.1]) on the \nobtained{} and \Jposterior{} metrics for the InfoBAX and MeanBAX strategies. InfoBAX generally exhibits greater robustness to noise relative to MeanBAX, characteristic of explorative acquisition strategies.}
\label{fig:noiseinf}
\end{figure}

\clearpage

%%% Each figure should be on its own page

\clearpage

\clearpage

%%% Each figure should be on its own page

\clearpage

\end{document}